\documentclass{elsart}
\usepackage[intlimits]{amsmath}
\usepackage{epsfig}

\newcommand{\Sc}{\scriptstyle}
\newcommand{\hpt}{\hspace{1pt}}
\newcommand{\bp}{\begin{pmatrix}}
\newcommand{\ep}{\end{pmatrix}}

\newcommand{\third}{{\textstyle\frac{1}{3}}}

\newcommand{\Slash}[1]{#1 \hspace{-.5em}/\hspace{.11em}} 
\newcommand{\fourint}[1]{\int\!\frac{d^4 #1}{(2\pi)^4}}

\newcommand{\vect}[1]{{\mbox{\boldmath $#1$}}}

\begin{document}

\begin{frontmatter}
{\small\noindent ADP-01-34/T466 \hfill FAU-TP3-01/8 \hfill UNITU-THEP-01/20}

\vspace{5mm}

\title{Relativistic Three-Quark Bound States in Separable Two-Quark 
Approximation}
  \author[Ade]{M.~Oettel},
  \author[Erl]{L.~von Smekal},
  \author[Tue]{R.~Alkofer}
  \address[Ade]{CSSM, University of Adelaide, 10 Pulteney St., Adelaide, 
                SA 5005,
                E-mail:~moettel@physics.adelaide.edu.au}
  \address[Erl]{Institut f\"ur Theoretische Physik III,
           Universit\"at Erlangen--N\"urnberg,
           Staudtstr.~7, 91058 Erlangen, Germany,
	   E-mail: smekal@theorie3.physik.uni-erlangen.de}
  \address[Tue]{Institut f\"ur Theoretische Physik,
           Universit\"at T\"ubingen,
	   Auf der Morgenstelle 14,
           72076 T\"ubingen, Germany,
	   E-mail: Reinhard.Alkofer@uni-tuebingen.de}
	   
\begin{abstract}

Baryons as relativistic bound states in 3-quark correlations are described 
by an effective Bethe-Salpeter equation when irreducible 3-quark
interactions are neglected and separable 2-quark correlations are
assumed. We present an efficient numerical method to calculate the
nucleon mass and its covariant wave function in this quantum field
theoretic quark-diquark model with quark-exchange interaction.   
Expanding the components of the spinorial wave function in terms of
Chebyshev polynomials, the four-dimensional integral equations are in
a first step reduced to a coupled set of one-dimensional ones.
This set of linear and homogeneous equations defines a generalised 
eigenvalue problem. Representing the eigenvector corresponding to the largest 
eigenvalue, the Chebyshev moments are then obtained by iteration.   
The nucleon mass is implicitly determined by the eigenvalue, 
and its covariant wave function is reconstructed from the moments
within the Chebyshev approximation.   

  \end{abstract}
\end{frontmatter}

\newpage

{\Large PROGRAM SUMMARY}

\textit{Title of program:} BSE




\textit{Computers:} Workstation DEC Alpha, LINUX PCs (AMD K7)

\textit{Operating system under which the program has been tested:} UNIX, Linux

\textit{Programming language used:} Fortran 90

\textit{Memory required to execute with typical data:} 10 MB

\textit{No. of bits in a word:} 32

\textit{Peripherals used:} standard output, disk

\textit{No.~of lines in distributed program, including test data, etc.:} 


\textit{Nature of physical problem:} 
Diquarks are introduced as separable correlations in the two-quark
Green's function to facilitate the description of baryons as relativistic
three-quark bound states. These states then emerge as solutions of
Bethe--Salpeter equations for quarks and diquarks that interact via
quark exchange.

\textit{Method of solution:} 
Chebyshev polynomials are used for an expansion of
the Bethe--Salpeter vertex and wave functions 
and an (approximative) representation of the interaction kernel and 
propagators.
The resulting set of coupled
one-dimen\-sional integral equations for the corresponding moments
of vertex and wave function is converted to a matrix equation
by employing Gaussian quadrature grid points. This equation is then solved 
iteratively for a certain test mass of the bound state. A bisection method
in the bound state mass variable is used to determine the correct 
mass corresponding to the chosen
quark and diquark parameters.

\textit{Restrictions on the complexity of the problem:} 
The separable quark-quark correlations are restricted to the 
leading covariants in the scalar and axialvector
(i.e.\ positive parity) channels.

\textit{Typical running time:} Depending on the parameters and the desired
accuracy between one and ten minutes  on 1.2 GHz AMD K7.

\vskip 20mm

\goodbreak

{\Large LONG WRITE-UP}

\section{The physical problem}

A Poincar{\'e}-invariant, quantum field theory based model of the nucleon is
required to describe the high-precision data of nucleon properties in
the several-GeV energy range. To this end, we here demonstrate how to
calculate (using certain approximations detailed below) the matrix elements 
of three quark operators $q$ between the vacuum 
$|\Omega \rangle $ and the nucleon bound state $| P_N \rangle$, the 
asymptotic momentum state with nucleon quantum numbers. 
This matrix element 
$\psi \sim  \langle \Omega | T(qqq) |P_N \rangle$
is the covariant Bethe--Salpeter
wave function of the nucleon which can be applied as input in covariant 
calculations of many observables such as the various form factors of the 
nucleon. In the following  sections we describe an approximation scheme
based on the Faddeev equations (see, {\it e.g.}, ref.\ \cite{Thomas:1977})
which makes use of the concept of diquarks as separable quark-quark
correlations \cite{Cahill:1989dx,Reinhardt:1990rw,Buck:1992wz,Ishii:1995bu,Zuckert:1997nu,Bender:1996bb,Hellstern:1997nv}.  
In this scheme, the covariant diquark-quark model, see 
refs.~\cite{Oettel:2000ig,Alkofer:2001ne,Alkofer:2000wg,Roberts:2000aa,Ahlig:2000qu,Oettel:2000jj,Oettel:1998bk,Bloch:2000rm}, 
we obtain a tractable set of equations for the components 
of the covariant nucleon wave function which can be solved without
non-relativistic reductions.

\subsection{The separable approximation to the relativistic 3-quark problem}

To summarize the relation between the covariant quark-diquark model
and the general, relativistic three-quark problem briefly in the following, 
we start from the six-quark Green function, 
\begin{equation}
G(x_i,y_i)=\langle0|
T\prod_{i=1}^3q(x_i)\bar{q}(y_i)|0\rangle .
\label{six}
\end{equation}  
The variables $x_i$ and $y_i$ not only represent the space-time 
coordinates of the quark fields, but also include their discrete
labels for color, spin, and flavor. The six-point function (\ref{six})
satisfies the Dyson equation 
\begin{equation}
G=G_0+G_0\circ K \circ G\, .
\label{G_Dyson}
\end{equation}
In this equation the disconnected 
six-point function $G_0$ describes the free propagation of 
three dressed quarks, and the three-quark scattering kernel $K$ contains 
all two-- and three-particle irreducible diagrams. The symbol 
``$\circ$'' in eq.~(\ref{G_Dyson}) denotes summation/integration over 
all independent internal coordinates and labels which defines a composition
law for linear maps on a suitable function space. Unless explicitly stated 
otherwise we will henceforth work in momentum space with Euclidean metric. 
It is thus not necessary to introduce different symbols for momentum and
coordinate space objects.

The nucleon as a three-particle bound state with mass $M$ manifests itself as
a pole  in the six-point function at $-P^2=M^2$ where  $P=p_1+p_2+p_3$ is the
total four-momentum of the three-quark system. One can thus
parameterize the six-point function in the vicinity of the pole as
\begin{equation}
 G(k_i,p_i) \sim \frac{\psi(k_1,k_2,k_3) 
\;\bar{\psi}(p_1,p_2,p_3)}{P^2+M^2} \; ,
 \label{bspole}
\end{equation}
where $\psi$ denotes the bound state wave-function. Substituting
this parameterization into the Dyson equation (\ref{G_Dyson})
and identifying the residuum contributions on both sides, one obtains 
the homogeneous bound state equation,
\begin{equation}
 \psi= G_0 \circ K \circ \psi  \qquad 
\Longleftrightarrow \qquad
G^{-1}\circ \psi=0\; .
 \label{3bpsi}
\end{equation}
Despite the simple appearance of the bound state equation in
this symbolic form, its direct solution is not feasible. Even in
principle, this could only be attempted once a model interaction
is specified to fix the contributions to the kernel  $K$ which 
can in general never be fully determined from first principles. 

The approximation scheme we present below essentially consists of two steps. 
Thereby, some of the contributions to the kernel, namely those due to
irreducible three-particle interactions, are neglected while an
explicit reference to the others is avoided by shifting the details of
the two-particle interactions into the parametrisations of the
relevant diquark correlations. 

The first step leads to the corresponding relativistic Faddeev
equations. Neglecting all contributions from irreducible three-quark 
interactions, the kernel $K$ can be written as a sum of three 
two-quark interaction kernels, 
\begin{equation} 
K\,=\,K_1+K_2+K_3 \; . 
\label{k2pi}
\end{equation} 
We adopt the notation that the subscript of $K_i$ refers to the spectator 
quark $q_i$. The $K_i = k_{qq} \otimes S_i^{-1}$ are defined in
3-quark space and thus contain an inverse quark propagator $S_i^{-1}$ 
acting on the subspace of the spectator quark.
The quark pair interacting via $k_{qq}$ in $K_i$ then is $(q_j,q_k)$ with
the three labels ($i,j,k$) being a cyclic permutation of $(1,2,3)$. 
These 2-quark kernels determine the interactions for  
disconnected scattering amplitudes $T_i = t_{qq} \otimes S_i^{-1}$ 
which describe the scattering between quark $j$ and $k$ (with
amplitude $t_{qq}$) and which are summed in the Dyson series for this
2-quark subspace as follows:  
\begin{equation} 
T_i= K_i  +  K_i \circ G_0 \circ T_i \; . 
\label{t_i}
\end{equation}
Introducing the Faddeev components of the 3-quark wave 
function $\psi$,
\begin{equation}
 \psi_i = G_0\circ K_i \circ \psi \; , \label{fad_comp}
\end{equation} 
one then has $\psi=\sum \psi_i$ by virtue of eqs.~(\ref{3bpsi},\ref{k2pi}).
Furthermore, eq.~(\ref{t_i}) can now be used to simplify
eq.~(\ref{3bpsi}) considerably yielding the coupled 
Faddeev equations,
\begin{equation}
 \psi_i  
   = G_0 \circ T_i\circ  (\psi_j+\psi_k) \;,
 \label{Faddeev}
\end{equation} 
for Faddeev components with pairwise distinct $i,j,k \in \{1,2,3\}$.

The assumption central to the second step, and the defining one of the
quark-diquark model, is the {\em separability} of the two-quark
scattering amplitude $t_{qq} $. This amplitude is thereby typically
parametrised by {\em scalar} and {\em axialvector} diquark
correlations to account for its most important separable contributions
at sufficiently small (absolute values of the complex Euclidean)
diquark momenta, 
\begin{eqnarray}
t_{qq}(k_j,k_k;p_j,p_k)&=& \chi^5((k_j-k_k)/2)\; 
D^{55}(k_j+k_k)\; \bar\chi^{5}((p_j-p_k)/2) + \nonumber \\
& &\chi^\mu((k_j-k_k)/2)\;D^{\mu\nu}(k_j+k_k)\; \bar\chi^{\nu}((p_j-p_k)/2)\; .
\label{sep_ass}
\end{eqnarray}
Here we employ simple particle-pole contributions 
to represent these diquark correlations. The diquark 
propagators are then given by the free scalar propagator and 
the Proca propagator for a spin-1 particle, respectively, 
\begin{eqnarray}
D^{55}(p) &=& -\frac{1}{p^2+m_{sc}^2} \; , 
 \label{Ds} \\
D^{\mu\nu}(p) &=& -\frac{1}
 {p^2+m_{ax}^2} \left( \delta^{\mu\nu}+ \frac{p^\mu p^\nu}{m_{ax}^2}
\right) \; . 
 \label{Da}
\end{eqnarray}  
The diquark-quark vertices $\chi$ in eq.~(\ref{sep_ass}) 
need to be antisymmetric with respect to the interchange of their
quark indices. Both diquarks belong to the 
antisymmetric color antitriplet. The scalar diquark vertex is an antisymmetric 
isospin singlet and is also antisymmetric in Dirac space while the 
axialvector diquark vertex is symmetric in both Dirac and isospin indices 
(triplet). We do not give their color and flavor structures explicitly
here, and we restrict the Dirac structures of both $\chi$ to the dominant 
ones which are given by  
\begin{eqnarray} 
\chi^5_{jk}(p)&=&g_s\;
(\gamma^5 C)_{jk}\; V_s(p) \; , \label{dqvertex_s} \\
\chi_{jk}^\mu(p)&=&g_a\; (\gamma^\mu C)_{jk}\; V_a(p)\; ,
\label{dqvertex_a} 
\end{eqnarray}
respectively, where $g_s$ and $g_a$ represent coupling constants which
are determined by the diquark normalizations, see below.
The scalar functions $V_{s, a}(p)$ parametrise the extensions of the two 
diquark-quark amplitudes in momentum space. Phenomenologically, a
dipole form for their dependence on the quark relative momenta has
proven successful, and we will assume equal widths for
both diquarks here, 
\begin{equation}
  V_s(p) = V_a(p)= \left(\frac{c_0^2}{c_0^2+p^2}\right)^2 \;, 
  \label{V}
\end{equation}
\nopagebreak
to reveal the general structures in a reasonably simple form 
in the following.

The diquark widths determine their couplings $g_s$ and $g_a$ by
normalization conditions derived from the Bethe-Salpeter
equation (\ref{t_i}) for the 2-quark scattering amplitude,
\begin{equation} 
t_{qq}= k_{qq}   +  k_{qq} \circ (S \otimes S) \circ t_{qq} \; . 
\label{t_qq}
\end{equation}
Inserting the ansatz (\ref{sep_ass}) with the diquark pole
contributions (\ref{Ds},\ref{Da}) into in the inhomogeneous equation 
(\ref{t_qq}) implicitly fixes the normalizations adopted for the
diquark amplitudes. These implicit conditions are most easily derived
from the derivative of eq. (\ref{t_qq}) w.r.t. the total
diquark momentum $P_d=k_j+k_k$, for details, see {\it e.g.}
refs.~\cite{Oettel:2000ig,Oettel:2000jj}. 
Under the mild additional assumption
that the 2-quark interaction kernel $k_{qq}$ be independent of $P_d$,
the normalization conditions for the scalar and axialvector diquark
amplitudes are,
\begin{eqnarray}
 \label{normsc}
 4 \, m_{sc}^2 \, &\stackrel{!}{=}& \,
        \bar\chi^5 \circ  \Big(  P_d\cdot\frac{\partial}{\partial P_d}   
 \big( S \otimes S \big) \Big) 
      \circ \chi^5 \; , \\ 
 \label{normax}
 12 \, m_{ax}^2 \, &\stackrel{!}{=}&\, 
   \bar\chi^\nu \circ  \Big(  P_d\cdot\frac{\partial}{\partial P_d}   
      \big( S \otimes S \big) \Big) \circ \chi^\mu  \; 
    \left( \delta^{\mu\nu} +  \frac{P_d^\mu P_d^\nu}{m_{ax}^2} \right) \; , 
\end{eqnarray}
respectively. The additional factor of $3$ in the last line arises
from the sum over the polarisation states of the spin-1 particle.

\subsection{The quark-diquark Bethe-Salpeter equation}

For identical particles, the three Faddeev components (\ref{fad_comp})
are obtained from a unique amplitude $\phi_{ijk}$ by cyclic
permutations of the indices and arguments labelling the individual quarks,
and the 3-quark amplitude is obtained by their sum or, equivalently,
their Faddeev wave function by $\psi = G_0 \circ \sum_{cyclic} \phi_{ijk} $. 
With the separable form for the two-quark scattering
amplitude $t_{qq}$ in eq.~(\ref{sep_ass}), 
the following {\em ansatz} is employed in the  
Faddeev equations~(\ref{Faddeev}): 
\begin{eqnarray}
 \phi_{ijk}(p_i,p_j,p_k) &=&   \label{fad_ans}\\
 && \hskip -2cm  
\chi^{\sf a }_{jk}(q)\;D^{\sf ab}\big( (1\!-\!\eta)P-p\big)
      \; \big(\Phi^{\sf b}(p,P) u(P)\big)_i\; , \quad 
      ({\sf a,b}=1\dots 5) \; . \nonumber
\end{eqnarray}
Herein, $u(P)$ is a positive-energy spinor for the nucleon state with
momentum $P=p_i+p_j+p_k$, $p=(1\!-\!\eta ) p_i - \eta
(p_j + p_k )$ is the relative momentum between quark and
diquark, and $q=(p_j-p_k)/2$ is the relative momentum of the 
two quarks within the diquarks. 
We have introduced a momentum partitioning parameter $\eta\in [0,1]$
which distributes the total momentum $P$ between quark and diquark.
An analogous parameter could also be introduced 
in a more general definiton of the relative momentum $q$ within the
diquarks as discussed in \cite{Oettel:2000jj}.
Translational invariance entails that observables, {\it i.e.} matrix
elements like the form factors etc., should not depend on such
momentum partitioning parameters in the covariant bound-state wave
functions. 
Furthermore, in eq.~(\ref{fad_ans}), $\Phi^{\sf a}$ is composed of 
Dirac matrices such that $\Phi^{5[\mu]} u$ is the most general 
effective [vector] spinor compatible with nucleon quantum numbers for
the Faddeev components. We call $\Phi^{\sf a}$ the nucleon 
Bethe-Salpeter vertex function with quark and diquark.

\begin{figure}
 \centerline{\epsfig{file=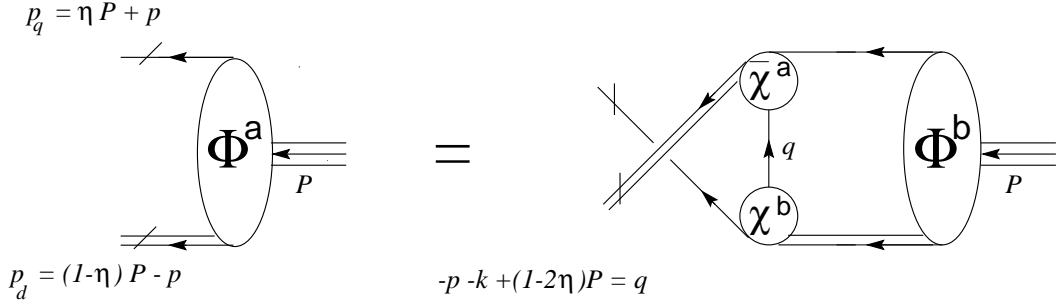,width=\linewidth}}
 \caption{The Bethe--Salpeter equation for the vertex function $\Phi$.}
 \label{bse_fig}
\end{figure}

The Faddeev equations (\ref{Faddeev}) with the ansatz~(\ref{fad_ans}) now 
reduce to a system of effective 2-particle Bethe-Salpeter equations 
for a quark-diquark state bound by repeated quark-exchanges.
Introducing an {\em eigenvalue} $\lambda $, for the lowest bound state
in the given channel, these equations have to be solved for the 
largest value of $\lambda \equiv \lambda(P^2)$ at a given $P^2$,  
\begin{eqnarray}
    \lambda(P^2)  \, \Phi^{\sf a}(p,P) &=& \, \fourint{k}\, 
    {\rm K}^{\sf ab}(p,k,P)\, \Psi^{\sf b}(k,P) \; , \quad 
  {\rm with} \label{bse}\\
 \Psi^{\sf a}(p,P)&=&S(\eta P+p)\,D^{\sf ab}\big((1\!-\!\eta)P-p\big)\,
      \Phi^{\sf b}(p,P) \;, \quad 
  {\rm and}   \label{bse1} \\
  {\rm K}^{\sf ab}(p,k,P) &=& -\frac{1}{2}
       \begin{pmatrix} -\frac{1}{g_s^2} \chi^5{\Sc (p_1) } \; S^T{\Sc (q) }\; \bar\chi^5{\Sc
  (p_2) } &  
     \frac{\sqrt{3}}{g_s^2}\; \chi^{\beta}{\Sc (p_1) }\; S^T{\Sc (q) }\;\bar\chi^5 {\Sc
  (p_2) } \\ 
    \frac{\sqrt{3}}{g_s^2}\;\chi^5{\Sc (p_1) }\; S^T{\Sc (q) }\;\bar\chi^{\alpha}{\Sc (p_2) }
    &  \frac{1}{g_s^2}\chi^{\beta}{\Sc (p_1) }\; S^T{\Sc (q) }\;\bar\chi^{\alpha}{\Sc (p_2) }
     \end{pmatrix} , \label{kern}
\end{eqnarray}
with the additional constraint that $\lambda(P^2) \!=\! 1/g_s^2 $ at $ P^2\! =
\! M_n^2$, thereby implicitly determining the nucleon mass.
We introduced the Bethe-Salpeter wave function $\Psi = {\rm
G}_0 \circ \Phi$ by attaching quark and diquark legs to the vertex function. 
With (\ref{bse1}) in eq.~(\ref{bse}) this determines a generalized
eigenvalue problem in which the vertex function $\Phi = {\rm K} \circ
{\rm G}_0 \circ \Phi$  
represents an eigenvector of the kernel K with respect to a
{\em metric} given by the disconnected quark-diquark propagator, 
\begin{equation}
  {\rm G}^{\sf ab}_0 =  S(\eta P+p)\,D^{\sf ab}\left( (1\!-\!\eta)P-p\right)\,
                       (2\pi)^4 \, \delta^4 (p-k) \; .
\label{qdqmet}
\end{equation}
The numerical factors in the elements of the 
quark-exchange kernel ${\rm K}^{\sf ab}$ arise from the
structure of the color and flavor couplings in the Faddeev equations.
Momentum conservation fixes the momentum of the exchanged quark,
\begin{equation}
  q   = -p-k+(1-2\eta)P \; ,
\end{equation}
and the relative momenta of the outgoing and the incoming quark-pair 
in $\chi(p_1)$ and $\bar\chi(p_2)$, respectively, are (see Fig.~\ref{bse_fig}),
\begin{eqnarray}
 p_1 &=& p+ k/2 -(1-3\eta)P/2 \; , \\
 p_2 &=& -k -p/2 + (1-3\eta)P/2 \; .
\end{eqnarray}
Since neither the eigenvalue $\lambda = 1/g_s^2$ nor the mass 
$M_n$ of the bound nucleon are allowed to depend on the momentum
partitioning parameter $\eta$, for
every solution  $\Psi(p,P;\eta_1)$ of the Bethe-Salpeter equation (\ref{bse})
there should exist a whole family of solutions of the form
$\Psi(p+(\eta_2-\eta_1)P,P;\eta_2)$.

\subsection{Decomposition of wave and vertex function}

Having defined the effective spinor of the spectator quark 
to be $\Phi^{\sf a} u$
in eq.~(\ref{fad_ans}) guarantees that the 3-quark wave function $\psi$ 
describes a spin-1/2 object. 
Furthermore, $\Phi^{\sf a}$ can most generally be chosen to    
be an eigenfunction of $\Lambda^+=1+\Slash{P}/(iM)$, the positive energy
projector, since $\Lambda^-u=(1-\Lambda^+)u=0$. Requiring positive parity 
for the Faddeev amplitudes leads to the additional condition
\begin{equation}
   \bp \gamma^4 \Phi^5(\tilde p, \tilde P) \gamma^4 \\ 
             \gamma^4 \tilde \Phi^\nu (\tilde p, \tilde P)  )\gamma^4 \ep
   \stackrel{!}{=}
   \bp \Phi^5(p,P) \\ - \Phi^\nu(p,P) \ep \; , \label{parity}
\end{equation}
with the parity operation on a 4-vector defined by $\tilde q=(-\vec q,q^4)^T$.

The positive-energy constraint and condition (\ref{parity}) greatly
restrict the number of independent components in the
vertex function. The scalar correlations $\Phi^5$ are described
by two components and the axialvector correlations by six
components,
\begin{equation} 
 \label{vex_N}
 \bp \Phi^5 (p,P)  \\ \Phi^\mu(p,P) \ep  =
 \bp
   \sum \limits_{i=1}^2 S_i(p^2,\hat p\cdot \hat P) \; {\mathcal S}_i(p,P)  \\[4mm]
   \sum \limits_{i=1}^6 A_i(p^2,\hat p\cdot \hat P) \; \gamma_5 {\mathcal A}_i^\mu(p,P) 
 \ep \; .
\end{equation}
The scalar functions $S_i$ and $A_i$  depend on the
two independent Lorentz invariants ($p^2$ and $p \cdot P$) that can be 
formed from the relative momentum $p$ and the on-shell nucleon momentum $P$. 
Normalized 4-vectors are introduced as $\hat p=p/|p|$, and for the
special case of the complex ({\it i.e.}, timelike) nucleon momentum we
adopt the convention that $\hat P= P/(iM)$, see below.

The Dirac components describing the scalar correlations
may be built out of
 $\Lambda^+$ and $\Slash{p} \Lambda^+$
and the Dirac part of the axialvector correlations 
can be constructed using the matrices
 $P^\mu \Lambda^+$, $P^\mu \Slash{p} \Lambda^+$, 
 $\gamma^\mu \Lambda^+$,   $\gamma^\mu \Slash{p} \Lambda^+$, 
 $p^\mu \Lambda^+$ and $p^\mu \Slash{p} \Lambda^+$.
Herein, for later convenience, we select a decomposition such that 
the scalar functions $S_i$ and $A_i$ decouple  
within the Dirac and Lorentz components of (\ref{vex_N}) 
in the nucleon rest-frame, 
\begin{equation}
P=(\vec 0,iM)^T \; , 
\label{rf1}
\end{equation} 
and with additionally choosing the spatial part of the relative
momentum to point in the $z$-direction (see sec.~\ref{subsec2.1}), 
\begin{equation}
p=(0,0,|p|\sin\psi,|p|\cos\psi)^T\; .
\label{rf2}
\end{equation}
A decomposition of the covariants achieving this turns out to be given by
\begin{equation}
 {\mathcal S}_i = \left\{ \begin{array}{l}{\mathcal S}_1=\Lambda^+ \\
                              {\mathcal S}_2=-\frac{i}{p}\, \Slash{p}_T\,\Lambda^+ 
                              \end{array} \right. \; , 
\qquad  {\mathcal A}^\mu_i=\left\{ \begin{array}{l}
       {\mathcal A}^\mu_1=-\frac{i}{p}\,\hat P^\mu \,\Slash{p}_T\,\Lambda^+ \\
       {\mathcal A}^\mu_2=\hat P^\mu \,\Lambda^+ \\
      {\mathcal A}^\mu_3=\hat p^\mu_T \,\hat \Slash{p}_T\, \Lambda^+ \\
     {\mathcal A}^\mu_4=\frac{i}{p}\, p^\mu_T \,\Lambda^+ \\
      {\mathcal A}^\mu_5=\gamma^\mu_T\, \Lambda^+ -{\mathcal A}^\mu_3 \\
     {\mathcal A}^\mu_6=
                  \frac{i}{p}\,\gamma^\mu_T\, \Slash{p}_T\, \Lambda^+-{\mathcal A}^\mu_4,
                \end{array}  \label{acov} \right. \; .
\end{equation}
Here and in the following, the subscript $_T$ denotes 
contraction with $\delta_{\mu\nu} - \hat P_\mu \hat P_\nu$  
to project 4-vectors onto the subspace transverse to the nucleon
momentum, {\it e.g.}, $p_T=p-\hat P (p \cdot \hat P)$. 
While the decomposition (\ref{acov}) is well suited for numerical processing, 
there exists an alternative decomposition of the Faddeev amplitude into 
components which for a nucleon at rest lead to
spin and orbital angular-momentum eigenstates
\cite{Oettel:2000ig}. The uniquely defined amplitudes 
in this partial wave decomposition are certain linear 
combinations of the functions $S_i$ and $A_i$ as defined by
eqs.~(\ref{vex_N},\ref{acov}) above.   

Since the constraints on the vertex function apply in the same way also
to the quark-diquark Bethe-Salpeter wave function $\Psi$ defined in
eq.\ (\ref{bse1}), its analogous decomposition reads,
\begin{equation} 
 \label{wex_N}
 \bp \Psi^5 (p,P)  \\ \Psi^\mu(p,P) \ep  =
 \bp
   \sum \limits_{i=1}^2 \hat S_i(p^2,\hat p\cdot \hat P) \; {\mathcal S}_i(p,P)  \\[4mm]
   \sum \limits_{i=1}^6 \hat A_i(p^2,\hat p\cdot \hat P) \; \gamma_5 {\mathcal A}_i^\mu (p,P)
 \ep \; .
\end{equation}
The coefficient functions $\hat S_i$,
$\hat A _i$ herein can be expressed
in terms of the $S_i, A_i$ that occur in the decomposition
(\ref{vex_N}) of the vertex function $\Phi$. 
The explicit expressions follow readily from the 
relation $\Psi = {\rm G}_0 \circ \Phi $, 
{\it c.f.}, eqs.\ (\ref{bse1},\ref{qdqmet}).

With the formalism being fully Poincar{\'e} covariant we are free to
choose the Lorentz frame that best suits our purpose of
solving the Bethe--Salpeter equations (\ref{bse}-\ref{kern}). 
As a particularly convenient choice 
we adopted the nucleon rest frame for the calculations presented
herein, with the total and relative momenta, $P$ and $p$,
as given in eqs.\ (\ref{rf1}) and (\ref{rf2}), respectively.

The  Bethe--Salpeter equations (\ref{bse}-\ref{kern}) 
are rewritten in terms of equations for the scalar functions
$S_1, \dots, A_6$. Their numerical solution described in the following 
section can be divided into two major steps:
\begin{itemize}
\item
Chebychev expansion of these scalar functions to account for their 
$\hat p\cdot \hat P$ dependences and
 Chebyshev polynomial approximation of the product of the free quark and
diquark propagators in the variable $\hat p\cdot \hat P$ and
of the quark exchange kernel in the two variables $\hat p\cdot \hat P$
and $\hat k\cdot \hat P$. 
\item
Solution by numerical iteration of the coupled system of
one-dimensional integral equations resulting for
the Chebychev moments of the scalar functions. 
\end{itemize} 
Note that the expansion in terms of Chebychev polynomials
employed in step two, though similar in nature, is not quite 
the same as a hyperspherical expansion. The Chebychev expansion
turns out to be extremely efficient here. Since typically only 
a few Chebychev polynomials are needed, see below, the numerical
effort is greatly reduced as compared to any attempt at calculating
the amplitude and the wave function directly on a two-dimensional grid.   

The underlying reason for this efficiency relates to the symmetries
of Bethe--Salpeter equations in certain limits. For a more detailed
discussion of the analogous method within the Wick--Cutkosky model
see, {\it e.g.\/}, section one of ref.\ \cite{Ahlig:1999qf} and the
references therein. The numerical solution of the 
ladder-approximate  Bethe--Salpeter equation in the massless as well
as the massive Wick--Cutkosky model has been studied extensively in
the literature, for a review see \cite{Nakanishi:1988hp} and for
a detailed numerical investigation see ref.~\cite{Nieuwenhuis:1996qx}. This
model describes the interactions of massive scalar particles by a
likewise scalar but possibly massless particle exchange.
In the case of a massless exchange-particle, 
the Bethe--Salpeter equation exhibits an $O(4)$ symmetry
\cite{Cutkosky:54}. By expanding the wave functions into hyperspherical
harmonics one finds that the zero orbital angular momentum states of the
Wick--Cutkosky model are essentially given by Chebyshev polynomials of the {\em
second} kind $U_n(z)$. These are related to the Chebyshev polynomials of the 
{\em first} kind $T_n(z)$, used in our expansions below, by 
\cite{Abramowitz:1965}
\begin{equation}
T_n(z)=U_n(z)-z U_{n-1}(z) \; , \qquad (n>0) \; . 
\end{equation} 
This thus exemplifies the close relation between the  
expansions of wave functions in terms of the $T_n$'s and those into
hyperspherical harmonics. 

The quark-diquark Bethe--Salpeter equations
(\ref{bse}-\ref{kern}), as derived from the relativistic Faddeev
equations under the separable assumption, result to be 
of a ladder-exchange type also. Here, the mixing of different angular
momentum components leads to additional structures, however. 
This is due to the spinor nature of the Bethe--Salpeter wave
function. Much the same as for ordinary Dirac spinors,
its lower components have different angular momentum than the upper ones
in the relativistic treatment of a spin-$\frac 12$ bound state. 
In order to obtain a closed system of equations, different
partial waves have to be taken into account. The expansion into 
{\em spinor} hyperspherical harmonics is nevertheless possible and
it convergences rapidly due to an approximate $O(4)$ symmetry
for fermion-boson Bethe--Salpeter equations of the kind under
consideration \cite{Oettel:2000jj,Oettel:1998bk}.
It is therefore understandable also for physical reasons that
the expansion in terms of Chebychev polynomials provides such an  
efficient numerical technique. 

To conclude this section, we discuss the kinematics which can be
employed to test the procedure and its limitations. 
In principle, the eigenvalue $\lambda=1/g_s^2$ should be
independent of the Mandelstam parameter $\eta$ (see the discussion after eq.\
(\ref{kern})). Due to singularities in the propagators, however, 
without considerable modifications to the 
numerical procedure, it is not possible to use
the whole interval [0,1] for such a test
\cite{Oettel:2000ig,Oettel:2000jj,Oettel:1998bk}. 
Rather, the range allowed for the momentum partitioning
parameter $\eta$ in the nucleon Bethe--Salpeter equation is restricted to
\begin{equation}
 \eta \in [ 1- m_{sc}/M_n, \, m_q/M_n ] \; ,
  \label{ebound1}
\end{equation}
if $m_{ax} \ge m_{sc} > m_q$ is assumed. Singularities in the exchange quark
propagator and the diquark vertices employing $n$-pole scalar functions $V$,
see eq.~(\ref{V}), lead to the additional bounds
\begin{eqnarray}
  \label{ebound2}
 \eta & \in & \left[\half(1-m_q/M_n), \, \half(1+m_q/M_n)\right] \; , \\
  \label{ebound3}
 \eta & \in & \left[\third(1-2\lambda_n/M_n), \, \third(1+2\lambda_n/M_n\right]
 \; .
\end{eqnarray}
Note that these bounds on the value of $\eta$ arise in the practical
calculations when performed as outlined above. Though an arbitrary
$\eta$ in [0,1] could be chosen in principle, additional residue terms
have to be included beyond these bounds in the Euclidean formulation 
to account for its proper connection with the underlying
Bethe--Salpeter equation in Minkowski space.

Well within the above bounds on the momentum routing, however,
pronounced plateaus can be observed in order to verify 
the $\eta$-independence of physical observables
\cite{Oettel:2000ig,Oettel:2000jj}. The extension of these
plateaus increases with increasing orders in the Chebyshev expansions.

As the values of $\eta$ approach the boundaries given in
eqs.~(\ref{ebound1}--\ref{ebound3}), however, strong variations 
occur due to the vicinity of the poles in the propagators 
with the effect that the convergence of their
Chebyshev expansion, {\it c.f.},  eq.~(\ref{eq1}) below,
is slowed down considerably. In order to achieve some sufficient
accuracy on the vertex function, larger and larger  
numbers of Chebyshev moments  $n_{\rm max}$ are then 
needed to approximate the components of the wave function $\psi$. 
The slower convergence of the wave function expansion under these
circumstances is also observed in the numerical solutions  
\cite{Oettel:2000ig,Oettel:2000jj,Oettel:1998bk}.


\section{Numerical method}

\subsection{Projection on scalar functions in nucleon rest frame}
\label{subsec2.1} 

In the nucleon rest frame (\ref{rf1},\ref{rf2}) we express the Euclidean
components of $p^\mu$ and $k^\mu$ (the latter being the integration variable)
in hyperspherical coordinates:
\begin{equation}
\bp k^1\\ k^2 \\ k^3 \\ k^4 \ep = |k| 
    \bp\sin\psi'\sin\theta'\sin\phi'\\\sin\psi'\sin\theta'\cos\phi'\\
             \sin\psi'\cos\theta'\\ \cos\psi' \ep \; ,
\quad \bp p^1\\ p^2\\ p^3\\ p^4 \ep =|p|
    \bp 0\\0\\\sin\psi\\ \cos\psi\ep \; . 
\end{equation}
To simplify the notation, we set $z=\cos\psi$, $z'=\cos\psi'$. 
In the chosen frame the vertex function decomposition on
the l.h.s of eq.~(\ref{bse}) simplifies to
\begin{equation} \label{pframe}
 \bp\Phi^5 \\ \Phi^4 \\ \Phi^3 \\ \Phi^2 \\ \Phi^1 \ep
 (p^2,z) = 
 \bp
  \bp S_1(p^2,z) &0\\ \sigma_3 \,\sqrt{1-z^2}S_2(p^2,z)&0 \ep \\
 \bp\sigma_3 \, \sqrt{1-z^2}A_1(p^2,z)&0 \\  A_2(p^2,z)&0 \ep \\
 \bp i\sigma_3 \, A_3(p^2,z)&0 \\ i \sqrt{1-z^2}A_4(p^2,z)&0\ep \\
 \bp i\sigma_2 \, A_5(p^2,z)&0 \\ -\sigma_1 \, \sqrt{1-z^2}A_6(p^2,z)&0\ep  \\
 \bp i\sigma_1 \,A_5(p^2,z)&0 \\ \sigma_2 \,\sqrt{1-z^2} A_6(p^2,z)&0\ep  \ep \;.
\end{equation}
By virtue of the positive-energy condition and our choice for 
the covariants in eqs.~(\ref{acov}) the vertex (and likewise
the wave function) consists of 2$\times$2--blocks in Dirac space
which define upper and lower components for the scalar and Lorentz components
of $\Phi$, respectively. The unknown scalar functions do not couple
within these blocks. Since we have chosen the most general decomposition,
the equations for the Lorentz components $\Phi^2$ and $\Phi^1$ are degenerate.

For convenience we will use generic functions $Y_i$ ($i=1,\dots, 8$)
related to the functions $S_i$ and $A_i$,
\begin{equation}
 S_{1,2} \rightarrow Y_{1,2} 
 ,\qquad A_{1\dots 6} \rightarrow Y_{3\dots 8}\; .
\end{equation}

The functions $\hat Y_i$ substituting the functions $\hat S_i,\hat A _i$ 
appearing in the wave function (\ref{wex_N}) are defined analogously.

Projection onto the scalar functions $Y_i$ and $\hat Y_i$ is 
now done by  inspecting the upper and lower components 
in the equations (\ref{bse1},\ref{bse})
for each Lorentz component. As explained before, the $Y_i$ and $\hat Y_i$
neatly decouple in upper and lower components.
We arrive at:
\begin{eqnarray}
  \label{bse2}
  \hat Y_i(p^2,z) &=& (g'_0)^{ij}(p^2,z) \;Y_j(p^2,z) \; ,  \\
   \label{bse3}
  Y_i (p^2,z)     &=& \fourint{k} \;(H')^{ij}(p^2,k^2,y',z,z') \;
                      \hat Y_i(k^2,z') \; .
\end{eqnarray}
As a result, we obtain
the propagator matrix $(g'_0)^{ij}$ and the modified 
quark exchange kernel matrix $(H')^{ij}$. 
The latter depends on the
possible scalar products between the vectors $k,p,P$. These scalar products can
be expressed as $k^2$, $p^2$, $z'=\hat k \cdot \hat P$, $z=\hat p \cdot \hat P$
and $y'=\hat \vect k \cdot  \hat \vect p$. We refrain from stating explicitely
the lengthy expressions for the quantities $(g'_0)^{ij}$ and $(H')^{ij}$ here.

\subsection{Chebyshev Approximation}

The approximation of a function by Chebyshev polynomials of the first
kind $T_n(z)$ is discussed in detail in ref.~\cite{Press:89}. We briefly
summarize the necessary formulae. Employing a convenient (albeit non-standard) 
normalization $T_0=1/\sqrt{2}$, the orthogonality relation reads,
\begin{equation}
 \int_{-1}^{1} \frac{T_n(z) T_m(z)}{\sqrt{1-z^2}}\;dz=\frac{\pi}{2}
   \delta_{nm} \; ,
  \label{ortho_T}
\end{equation}
Let $\{z_k,k=1\dots K\}$ denote the zeros of $T_{K}$. 
The discrete orthogonality relation is
\begin{equation}
 \frac{2}{K}\sum_{k=1}^{K} T_m(z_k) T_n (z_k) = \delta_{mn} \; .
  \label{ortho_T_disc}
\end{equation}
We  approximate a function $F(p^2,p\cdot P)$ by a finite sum of
$T$'s in the variable $z= \hat p \cdot \hat P$:
\begin{eqnarray}
 \label{Cheby_deco}
 F(p^2,p\cdot P)&=&\sum_{n=0}^{n_{\rm max}} i^n F^n(p^2) T_n(z)\; , \\
 \label{Cheby_mom}
 F^n(p^2) &=& (-i)^n \sum_{k=1}^{n_{\rm max}+1} T_n(z_k) F(p^2,z_k)\; .
 \label{approx}
\end{eqnarray}
As before, the $\{z_k\}$ are the zeros of  $T_{n_{\rm max}+1}$.
Note that for a finite $n_{\rm max}$ 
the such defined Chebyshev moments $F^n(p^2)$ 
are {\em not} identical to the projections of eq.~(\ref{Cheby_deco}) 
using eq.~(\ref{ortho_T}), rather eqs.~(\ref{Cheby_deco},\ref{Cheby_mom})
are very close to the approximation of $F$ by the {\em minimax polynomial},
which (among all polynomials of the same degree) has the smallest maximum
deviation from the true function \cite{Press:89}.
For $z=z_k$ and/or $n_{\rm max} \to \infty$, projection and approximation
are of course identical, courtesy of eq.~(\ref{ortho_T_disc}). 

We use this method to approximate propagators and the exchange kernel in
 eqs.~(\ref{bse2},\ref{bse3}).
Expanding the amplitudes of wave and vertex function as
\begin{eqnarray}
 \label{cheby-v}
  Y_i(p^2,z) &=& \sum_{m=0}^{m_{\rm max}} i^m Y_i^m(p^2)\;T_m(z) \; ,\\
  \hat Y_i(p^2,z) &=& \sum_{n=0}^{n_{\rm max}} i^n
      \hat Y_i^n(p^2)\;T_n(z)\; ,  \label{cheby-w}
\end{eqnarray}
and applying eqs.~(\ref{Cheby_deco},\ref{Cheby_mom}) 
to $(g_0')^{ij}$ in eq.~(\ref{bse2}),
we obtain the matrix equation
\begin{eqnarray}
  \hat Y_i^n(p^2) &=& (g_0)^{ij,nm} (p^2)\; Y_j^m(p^2) \; ,
 \label{eq1} \\
  (g_0)^{ij,nm} (p^2) &=& \frac{2}{p_{\rm max}+1} \;\sum_{k=1}^{p_{\rm max}}
   \;(g_0')^{ij}(p^2,z_k)\;i^{j-i}\; T_n(z_k)\, T_m(z_k) \label{g0} \; .
\end{eqnarray}
The projection onto the $Y_i^n$ was done using the orthogonality
relation (\ref{ortho_T}).
In practice, we choose $p_{\rm max}=n_{\rm max}+m_{\rm max}$ for
a reliable approximation of the propagator matrix. The elements of the propagator matrix $(g_0)^{ij,nm}$ are all real due to the
explicit phase factor $i^n$ in the Chebyshev expansions
(\ref{cheby-v},\ref{cheby-w}).

The Chebyshev approximation in the variables $z$ {\em and} $z'$ is now
applied to the matrix $(H')^{ij}$ in eq.~(\ref{bse3}):
\begin{eqnarray}
  (H'')^{ij}& =& (1-z^{\prime 2})(H')^{ij} = \sum_{s=1}^{m_{\rm max}+1}
    \sum_{t=1}^{n_{\rm max}+1} c^{ij,st} \; T_{s-1}(z) T_{t-1}(z') \; .
 \label{cap1} \\
  c^{ij,st} &=&
    { \frac{2}{(m_{\rm max}+1)} \frac{2}{(n_{\rm max}+1)}}
   \sum_{u=1}^{m_{\rm max}+1} \sum_{v=1}^{n_{\rm max}+1}
   \, T_{s-1}(z_u)T_{t-1}(z'_v) \nonumber \\
 & &   \times \; (H'')^{ij}(p^2,k^2,y',z_u,z'_v) \; .
 \label{cap2}
\end{eqnarray}
The $\{z_u\}$ and $\{z'_v\}$ are the  zeros of the Chebyshev polynomial 
$T_{m_{\rm max}+1}(z)$ and $T_{n_{\rm max}+1}(z')$, respectively. 
We insert this expression
into eq.~(\ref{bse3}), as well as the expansions (\ref{cheby-v},\ref{cheby-w})
for wave and vertex function. After projecting onto the $Y^m_i$
the integrations over $z$ and $z'$ on the r.h.s. are done 
using the orthogonality
relation (\ref{ortho_T}) and we obtain:
\begin{eqnarray}
 Y^m_i(p^2)&=& \frac{1}{(m_{\rm max}+1)(n_{\rm max}+1)}
  \int_{0}^\infty \frac{k^3}{4\pi^2} d|k| \int_{-1}^{1} dy'
 \sum_{u=1}^{m_{\rm max}+1} \sum_{v=1}^{n_{\rm max}+1}\;i^{n-m}
 \nonumber \\
  & & \times \;T_{m}(z_u)T_{n}(z'_v) \; (H'')^{ij}(p^2,k^2,y',z_u,z'_v)
   \; \hat Y^n_j (k^2)
 \; .
\end{eqnarray}
Here the sum runs also over the label $j$ and the Chebyshev moment label $n$.

We have now  succeeded to transform the original 4-dimensional integral
equation into a system of coupled one-dimensional equations. In summary, the
system reads,
\begin{eqnarray}
 \label{BS8}
 \hat Y_i^n(p^2) &=& (g_0)^{ij,nm} (p^2)\; Y_j^m(p^2) \; , \\
 \label{BS8a}
  Y_i^m(p^2)     &=& \int_{0}^\infty d|k| \;H^{ij,mn}(k^2,p^2)\;
                       \hat Y^n_j (k^2) \; ,
\end{eqnarray}
where indices appearing twice are summed over. Furthermore, the definition
\begin{eqnarray}
  H^{ij,mn}(k^2,p^2)&=& \frac{1}{(m_{\rm max}+1)(n_{\rm max}+1)}
    \frac{k^3}{4\pi^2} \int_{-1}^{1} dy'  \sum_{u=1}^{m_{\rm max}+1}
       \sum_{v=1}^{n_{\rm max}+1}\;i^{n-m} \nonumber \\
    \label{kern_final}
    & &   \times \;    T_{m}(z_u)T_{n}(z'_v) \; (H'')^{ij}(p^2,k^2,y',z_u,z'_v)\; 
\end{eqnarray}
is used. 

\subsection{Solving the coupled one-dimensional integral equations}

The integration variable in eq.\ (\ref{BS8a}) is the absolute value of the 
four momentum $k$. It will be discretized on a mesh with $n_{|k|}$ points, with
typically $n_{|k|}= 20,\dots, 50$, and the integration is performed as a
Gaussian quadrature. In the actual version of the program, the integration
domain is first mapped onto the interval [-1,1] with the help of the
transformation $x=(|k|-1)/(|k|+1)$ and then a Gauss--Legendre integration is
used.

After this step, the problem has become equivalent to finding the largest
eigenvalue of a matrix equation.
To this end we employ an iterative method: Some initial values
for the vertex functions, {\it i.e.} for the moments  $Y_i^m$, are chosen, and
eq.\ (\ref{BS8}) is used to calculate the corresponding expression for the 
wave function, {\it i.e.} for the moments $\hat Y_i^n$. Eq.\ (\ref{BS8a})
provides the moments $Y_i^m$ in the next iteration step. If $Y_1^0(p^2)$
deviates more than a given accuracy (provided in the input file) for some
momentum on the grid the updated moment $Y_1^0(p^2)$ at the grid point with the 
largest deviation is used to update the coupling constant $g_s$. This value and
renormalized functions $Y_i^m$ are then the starting point for the next
iteration step. Requiring an accuracy of the order $10^{-9}$ convergence is
usually reached after 20 - 30 iteration steps.

The correct nucleon mass corresponds to the eigenvalue $\lambda=1/g_s^2$.
To determine it, in a first step eigenvalues $\lambda(M_1)$ and $\lambda(M_2)$
are calculated with $M_1={\rm max}[m_q,m_{sc},m_{ax}]$ and
$M_2={\rm min}[m_q/\eta, m_{sc}/(1-\eta), m_{ax}/(1-\eta)]$. These starting
values are motivated by the observation that for physically reasonable 
parameters the nucleon mass should be larger than the quark and diquark 
masses, and, for a given choice of $\eta$, $M_2$ represents the upper limit
beyond which singularities are encountered in the quark and diquark
propagators. Now the nucleon mass $M$ for which $\lambda(M)=1/g_s^2$ holds
is determined by linear bisection in the next step and by quadratic 
interpolation in all following steps. Typically 5 - 8 steps are needed
to arrive at an accuracy of the order of $10^{-5}$. 
During each step the integral equation has to be solved anew for its largest
eigenvalue, thus illustrating the need for a fast algorithm which is provided
here.

As the system under consideration is a linear integral equation, the
normalization for the functions $Y_i^m$ and $\hat Y_i^n$ stays undetermined.
Without loss of generality we fix the value of the lowest Chebychev moment of
the first upper component of the vertex function at the smallest absolute value
on the momentum grid,
\begin{equation}
Y_1^0(p^2_{\rm min} \approx 0) = 1,
\end{equation}
and renormalize accordingly all functions $Y_i^m$ and $\hat Y_i^n$ in the 
output routine. 
These vertex and wave functions 
are then the final result of the algorithm.


\section{Numerical Results}

\begin{table}
 \begin{center}
  \begin{tabular}{ccccccccc} \hline \hline \\
  Set & $\eta$
  & $m_q$& $m_{sc}$& $m_{ax}$ & $c_0$ & $g_s$ & $g_a$  & $M$ \\
  & & [GeV]& [GeV]&   [GeV]&   [GeV]&                     &  & [GeV]
  \\[2mm] \hline
   I  & 0.36 & 0.360& 0.6255   & 0.6840  &   0.95 &     9.29  & 6.97    & 0.939\\
   II & 0.40 & 0.425& 0.5977   & 0.8314 &   0.53 &    22.10 & 6.37    & 0.939
 \\ \\ \hline \hline
  \end{tabular}
 \end{center}
 \caption{The two parameter sets together with the values of 
couplings and the bound mass $M$ that arise for these sets. The 
maximal number of Chebyshev polynomials for vertex and wave function
is given by $m_{\rm max}=8$ and $n_{\rm max}=12$ and the number of
momentum grid points is $n_{|k|}=40$. The $y'$ integration was done using
Gaussian quadrature with $n_y=32$ grid points.}
\label{pars}
\end{table}

To keep this article reasonably self-contained we will present numerical
results only for two example sets of parameters. Quite a number of results
obtained with this program can be found in refs.\ 
\cite{Oettel:2000ig,Ahlig:2000qu,Oettel:2000jj,Oettel:1998bk}. Altogether there
are four parameters: the quark mass $m_q$, the diquark masses $m_{sc}$ and
$m_{ax}$, and the width $c_0$ of the dipole-shaped diquark-quark vertex 
(\ref{V}). 
Diquark masses and the width determine the coupling constants $g_a$ and 
$g_s$. 
Note that in actual applications $g_a$ and therefore $m_{ax}$ is
determined by fitting the mass of the $\Delta$ baryon, see {\it e.g.} ref.\
\cite{Oettel:2000ig} for more details. The equation for the $\Delta$ baryon is
solved by an algorithm completely analogous to the one presented here. 
The results for two characteristic parameter sets 
are given in table~\ref{pars}.

\begin{figure}
 \begin{center}
   \epsfig{file=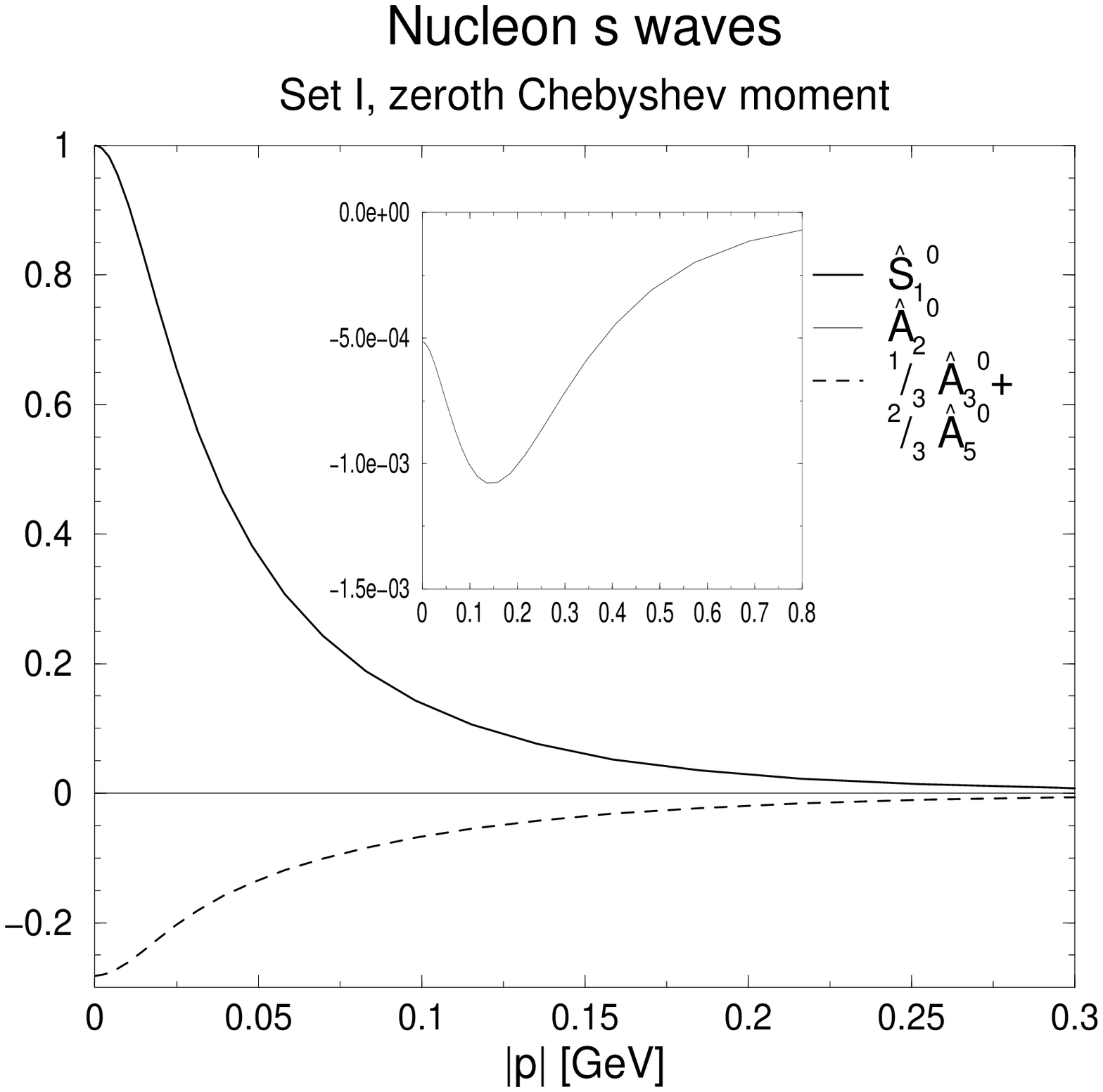,width=0.48\linewidth}
   \epsfig{file=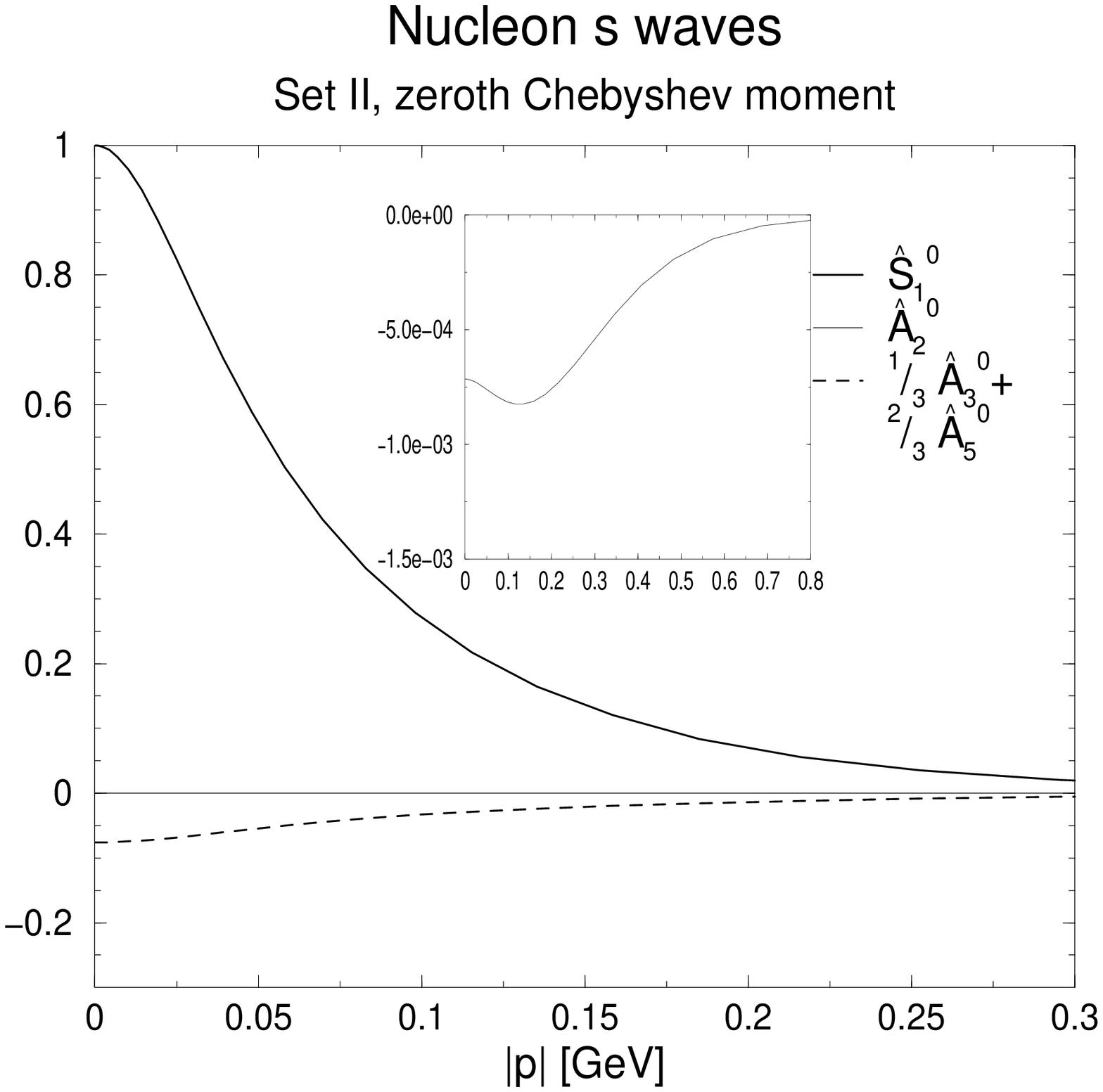,width=0.48\linewidth}
 \end{center}
 \caption{The leading Chebyshev moments of the
   functions $\hat S_1$, $\hat A_2$ and $(1/3) \hat A_3 + (2/3)\hat A_5$ 
   related to the
   nucleon $s$ waves. All functions are normalized by the condition
   $\hat S_1^0=1$.}
 \label{s_wfig}
\end{figure}

Although the wave and the vertex function are no physical observables they do
enter observable matrix elements (see {\it e.g.} refs.\
\cite{Ahlig:2000qu,Oettel:2000jj,Oettel:1998bk}) and therefore the strengths of
the single components give a hint on their effect on observables. We have
plotted the leading Chebyshev moments of the scalar functions describing the
nucleon $s$ waves in figure \ref{s_wfig}. These are $\hat S_1^0$ and 
$(1/3) \hat A_3^0 + (2/3) \hat A_5^0$, 
describing the $s$ waves for scalar and axialvector diquark and $\hat
A_2^0$ that is connected with the virtual time component of the latter. We see
for both sets that the functions $\hat A_2^0$ are suppressed by a factor of
$10^3$ compared to the dominating $\hat S_1^0$. The strength of the other $s$
wave associated with the axialvector diquark is roughly proportional to the
ratio $g_a/g_s$ for the respective parameter set.

In figure \ref{i} we have plotted the $0^{th}$, $2^{nd}$ and $4^{th}$ Chebychev
moments (the odd ones have been left out for clarity of the presentation) 
of  the linear combinations of the functions $\hat
S_i$ and $\hat A_i$ which belong to the eigenfunctions to the
(3 quark) spin operator and the operator of angular momentum between quark
and diquark \cite{Oettel:1998bk}.
One sees easily that in the nucleon rest frame higher
Chebychev moments become unimportant very fast.

\newlength{\figurewidtha}
\setlength{\figurewidtha}{0.37\linewidth}
\begin{figure}[p]
 \begin{center}
   \epsfig{file=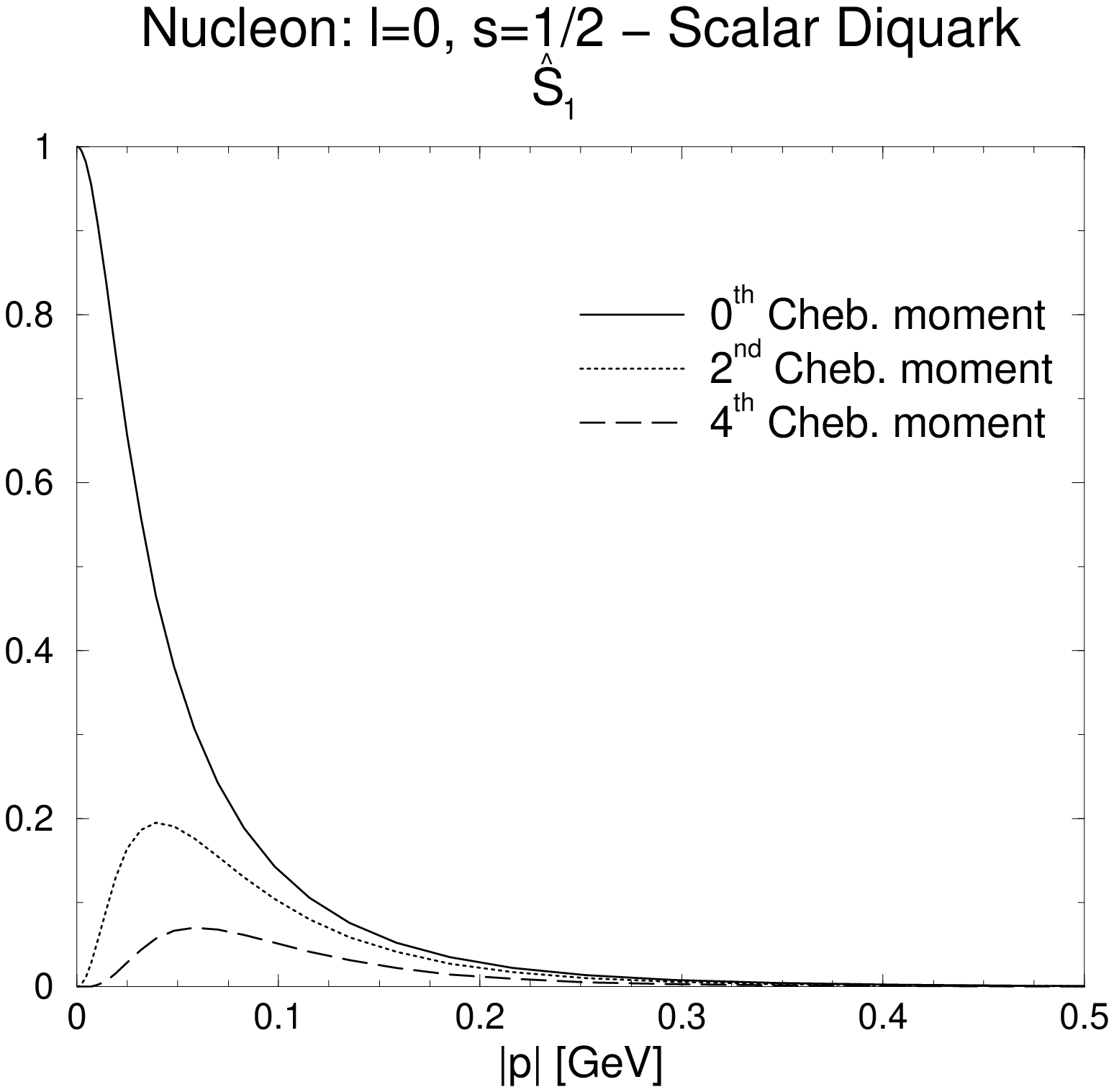,width=\figurewidtha} \hspace{5mm}
   \epsfig{file=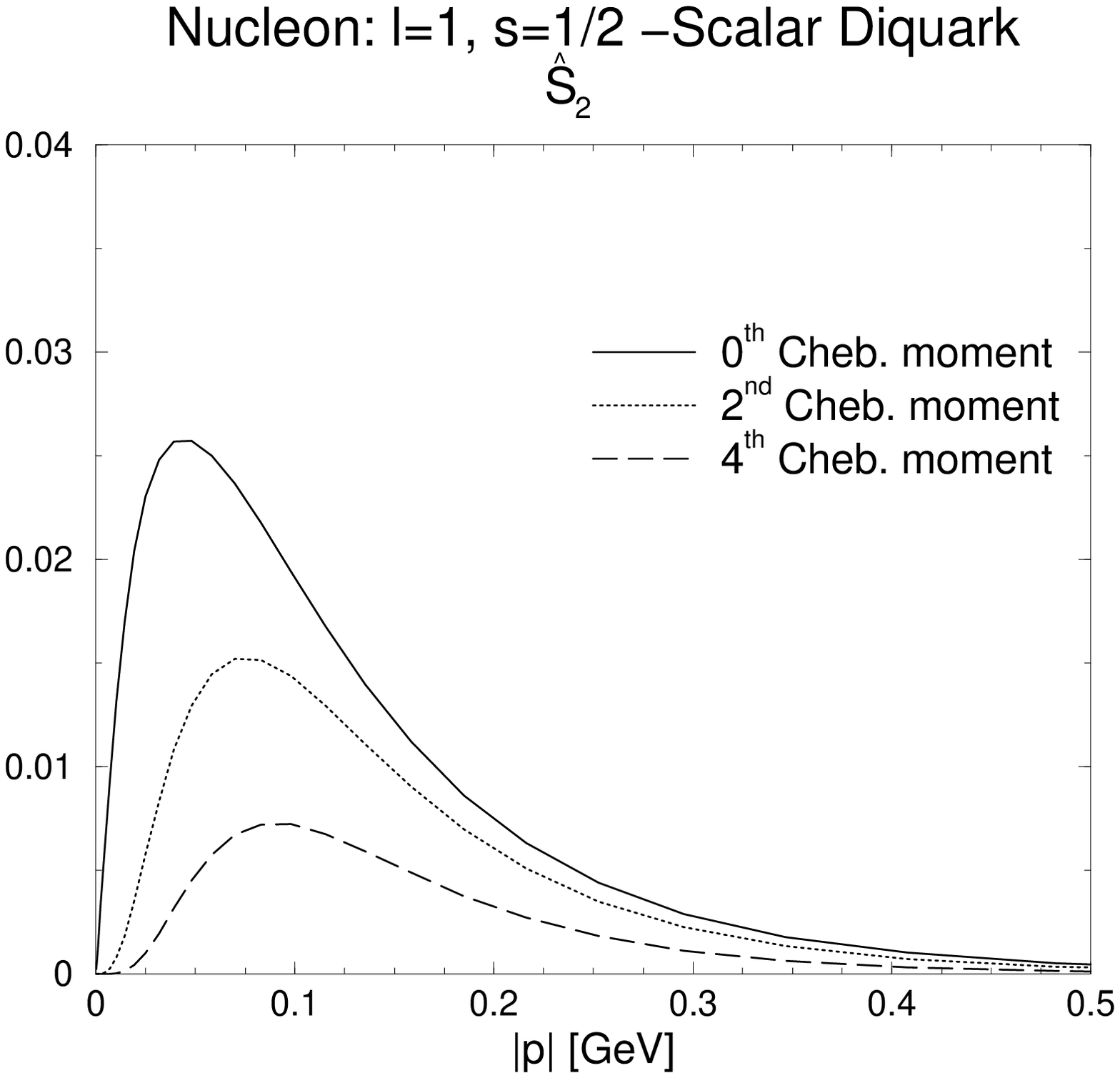,width=\figurewidtha} \hspace{5mm}
   \epsfig{file=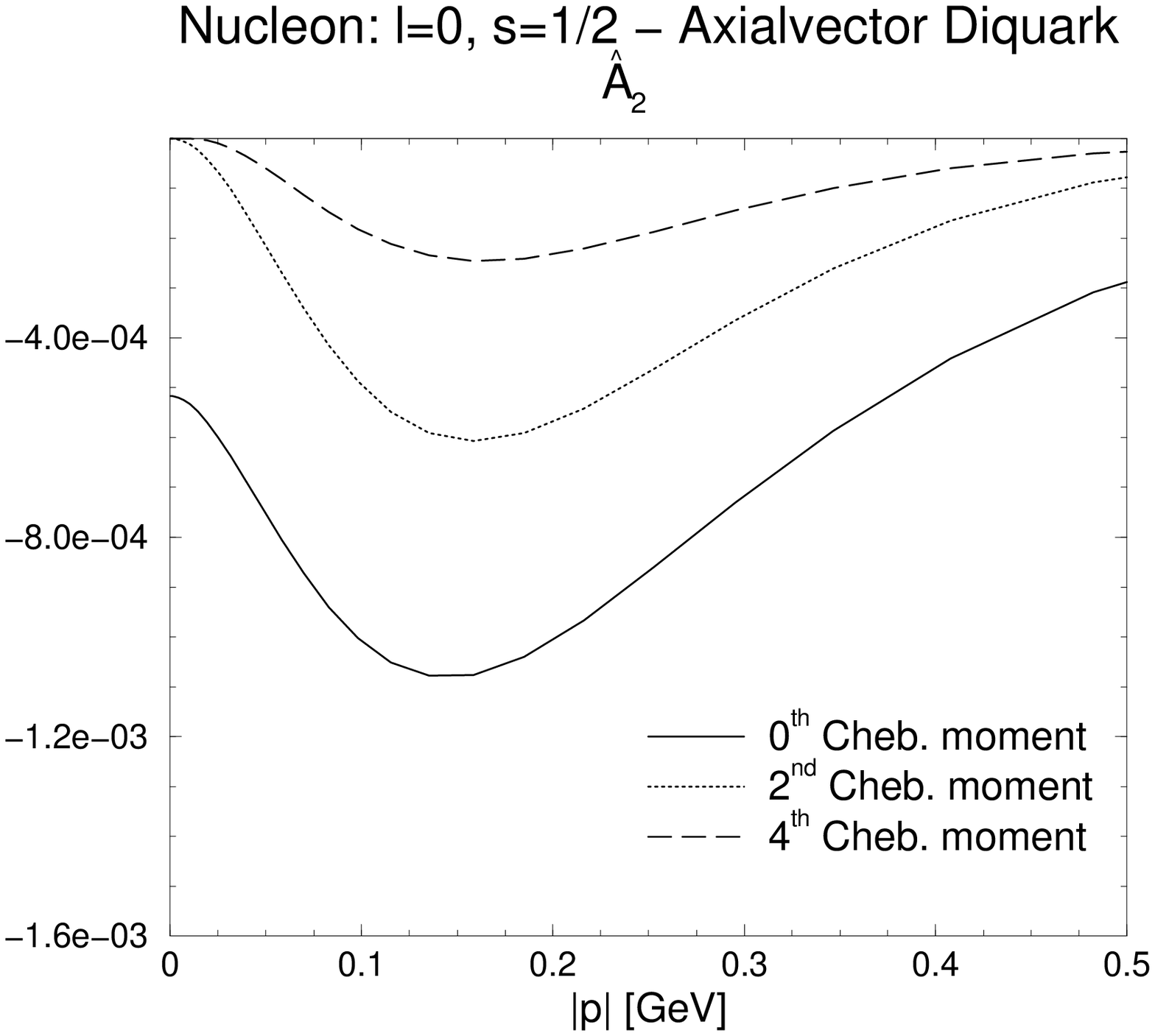,width=\figurewidtha} \hspace{5mm}
   \epsfig{file=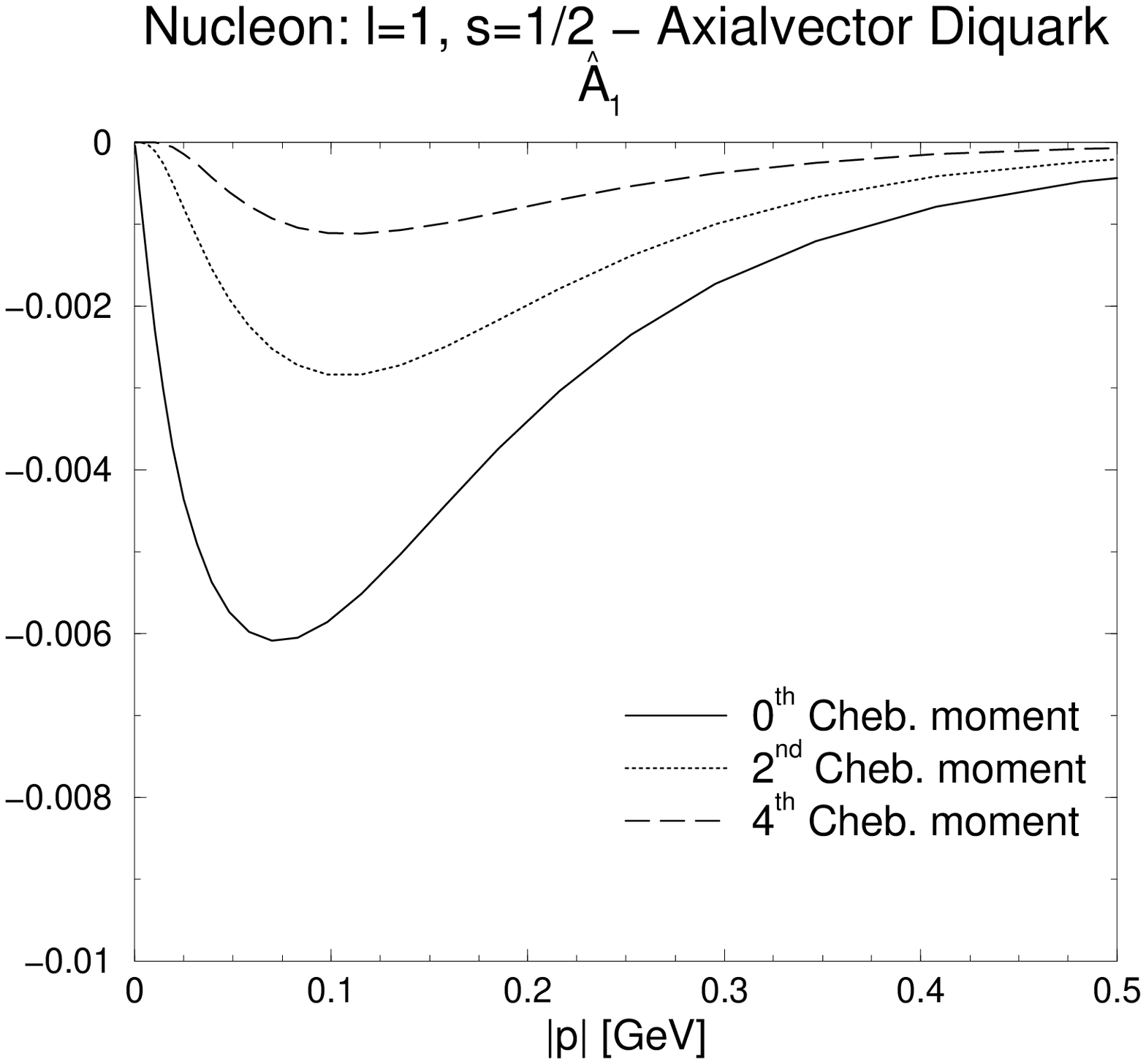,width=\figurewidtha} \hspace{5mm}
   \epsfig{file=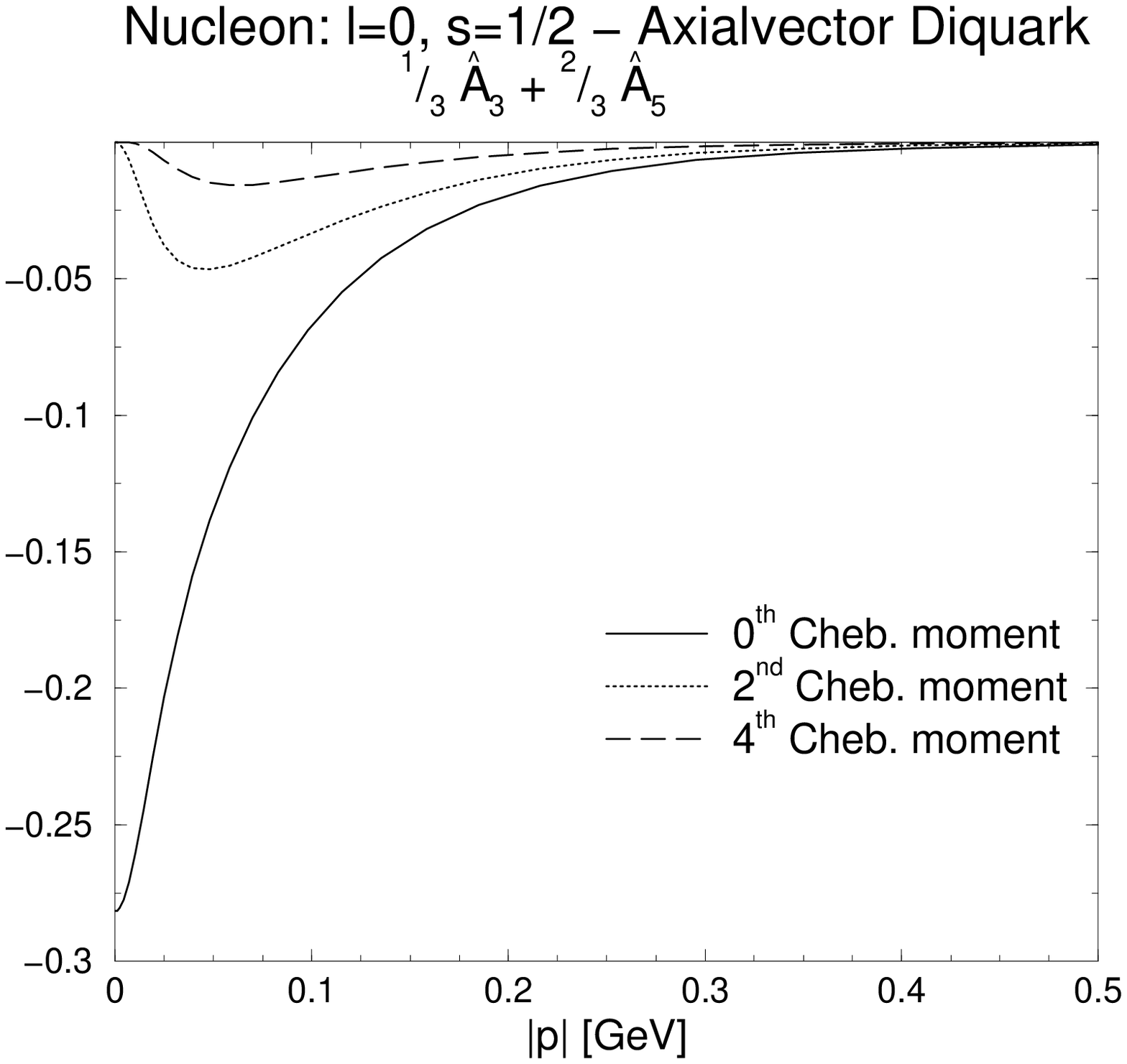,width=\figurewidtha} \hspace{5mm}
   \epsfig{file=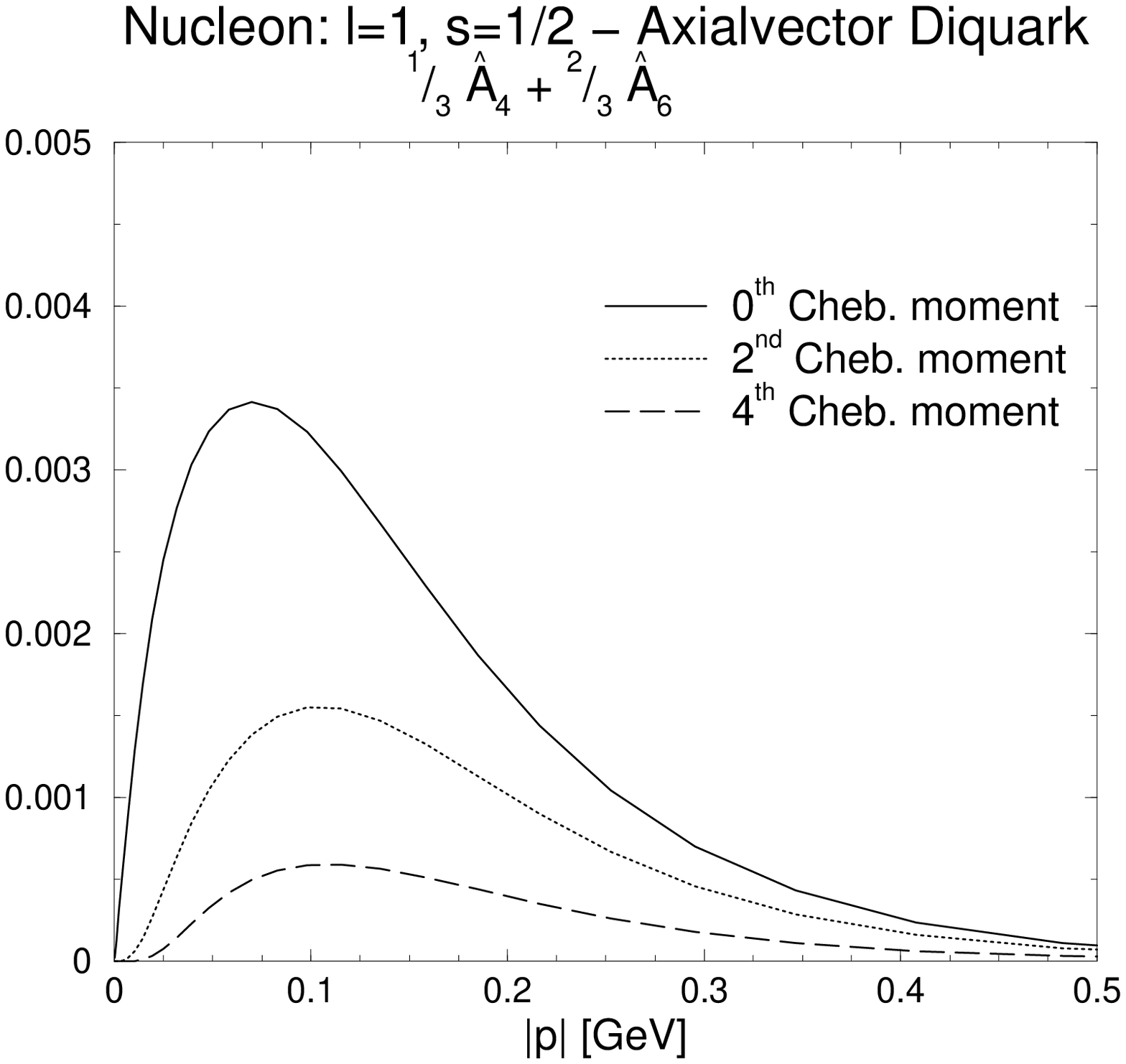,width=\figurewidtha} \hspace{5mm}
   \epsfig{file=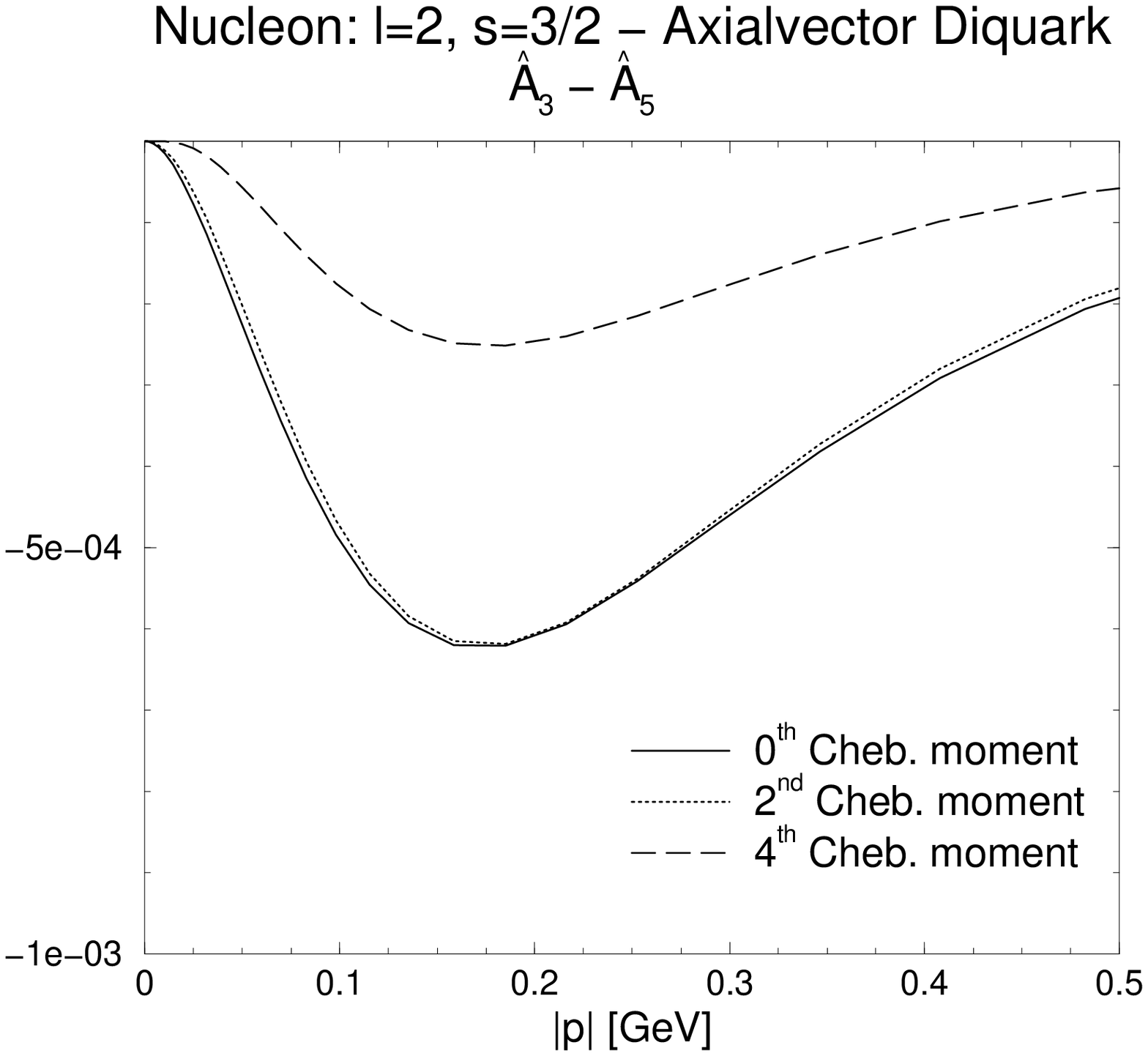,width=\figurewidtha} \hspace{5mm}
   \epsfig{file=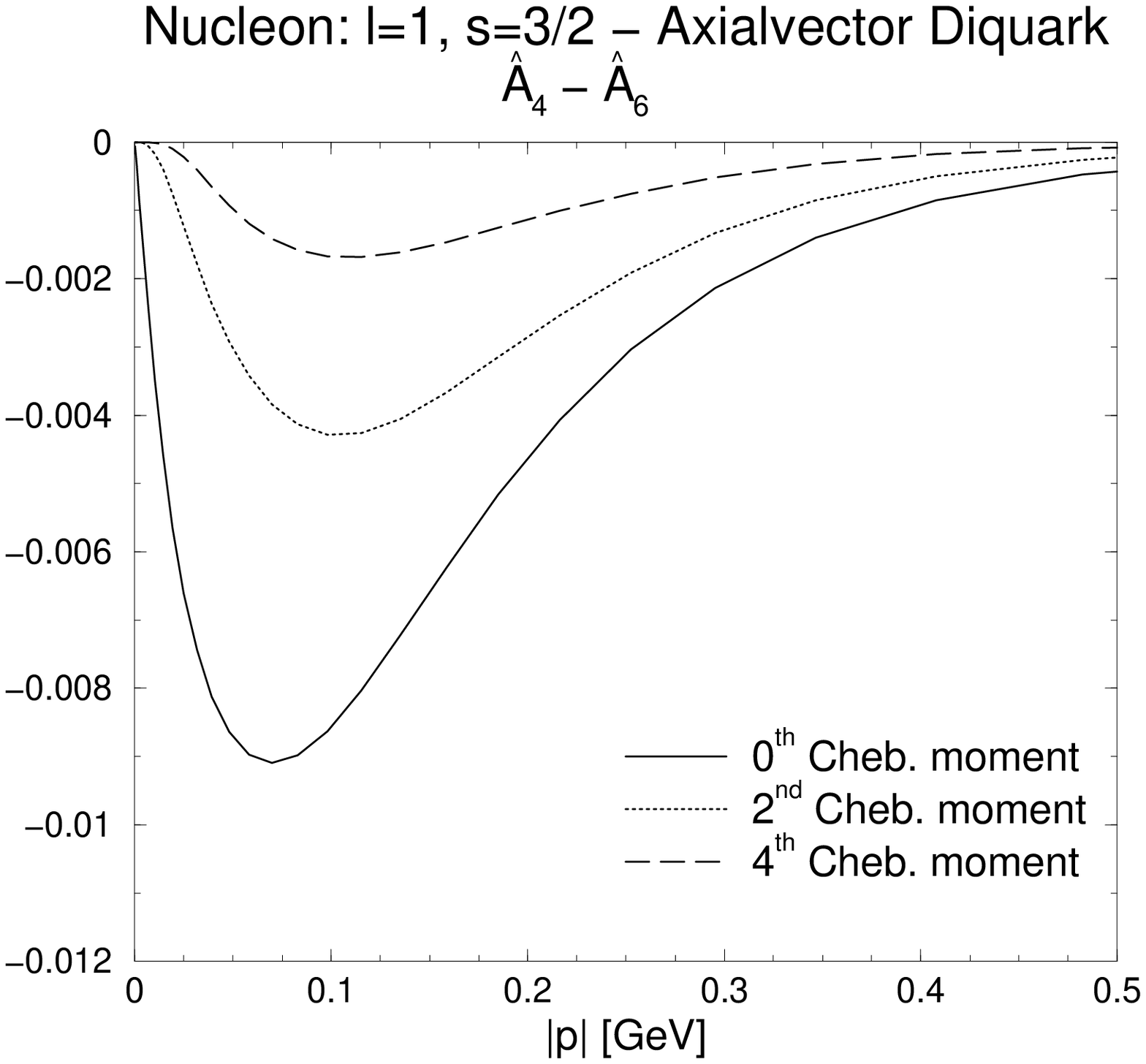,width=\figurewidtha}
 \end{center}
 \caption{Chebyshev moments of the scalar functions describing the
  strengths of quark--diquark partial waves with quantum number
  {\sf s} for the total quark--diquark spin and quantum number {\sf l}
  for the orbital angular momentum.
   }
 \label{i}
\end{figure}

\section{Summary and Applications}

In summary, we have succeeded in developing an efficient algorithm for the
calculation of the nucleon amplitudes in the seperable approximation to the 
Poincar{\'e} covariant three-quark Faddeev equation. To this end we  employ an
approximate $O(4)$ symmetry of the resulting effective diquark-quark
Bethe--Salpeter equation by using a Chebychev expansion. This allows to
transform the full 4-dimensional problem into a manageable form. 

The chosen expansion in Dirac space into partial waves has been done
covariantly and the results for the partial wave strengths identify the leading
and subleading  components. The Chebyshev expansion for the scalar functions
which describe the partial wave strengths has proved to be quite efficient and
accurate, furthermore the expansion has been done for a Lorentz invariant
variable, with the consequence that the boost of the wave and vertex function
solutions is rendered feasible.

The latter point is of utter importance because the applications of the
presented results to calculate physical observables require to Lorentz boost
the Bethe--Salpeter wave function. Please note also that in Lorentz frames 
where the nucleon is moving higher Chebychev moments will become increasingly
important with increasing nucleon momentum. Nevertheless, based on the
presented algorithm nucleon form factors have been calculated up to several GeV
of transfered momentum \cite{Oettel:2000jj}, and even production processes have
been studied for a similar energy range \cite{Ahlig:2000qu}. 

Given the fact that the required CPU time to run this program is very moderate
even on PCs, and that the underlying Poincar{\'e} invariant nucleon model has a
much broader application range than non-relativistic models we are confident
that the presented model and the related algorithm will find wide-spread use in
future investigations of nucleon properties.


\section{Description of the Program}

\subsection{The main program}

Before the main program we have included a {\bf module global} to provide a
global definition of the input parameters. Furthermore, the integration points
and weights are defined globally.

The main program calls the subroutine {\bf read\_parameters}, allocates memory
space for the used fields, and calls then the subroutine {\bf int\_weights}
three times determining thereby the integration points and weights for the one
angle and the one momentum integration. Note that the momenta $pp\in [0,\infty
]$ are mapped to the interval [0,1] for a Gauss--Legendre quadrature of the 
momentum integration. 

The subroutine {\bf dqnorms} is called to calculate $g_s$ and $g_a$
as prescribed by eqs.~(\ref{normsc},\ref{normax}).

Then the iteration loop for the nucleon mass is started with two initial
values for the 
bound state mass $M$. During each iteration step
the main algorithm, given in the subroutine {\bf bsemain} is called
which calculates
the vertex function, the wave function and the coupling 
$g_s=1/\sqrt{\lambda}$ which corresponds to the mass $M$. 
The mass is updated by bisection until satisfactory agreement between
$1/\sqrt{\lambda}$ and the scalar diquark normalization is reached.

Calling the 
subroutine {\bf output} concludes the main program.

\subsection{Input and Output}

Reading the input and writing out the vertex function, the wave function
and the coupling $g_s$is defered to the subroutines {\bf read\_parameters} and
{\bf output}. The format in which the parameters are read in and the results
are written out is obvious from the appended sample input and sample output.

\subsection{The subroutines ``bsemain'' and ``bse\_kernel''}

The first action to be taken in the subroutine {\bf bsemain} is to call 
the subroutine {\bf bse\_kernel} to generate the kernel of the equation
as given in eq.\ (\ref{kern_final}). This includes the integration over 
$y'$ and the summation over the zeros of the Chebychev polynomials. A complex 
auxiliary array $s$ is introduced whose real part, denoted $r$, is then used
to provide the explicit expression for the kernel.

Next,  the propagator matrix $(g_0)^{ij,mn}$ is generated,
via the intermediary steps of specifying the numerator of the matrix
$(g_0')^{ij}$ and the application of eq.~(\ref{g0}).

Then the Chebychev moments $Y_i^m$ are initialized, all by the same
function $pe^{-p}$. The iteration is implemented by a DO WHILE loop. 
In the iteration loop, first the Chebychev expansion of the propagators
(denominators and numerators) and in a subsequent step the one of the product
of quark and diquark propagators is determined. This allows to calculate the
functions $\hat Y_i^n$, {\it i.e.} the wave function $\Psi$, from eq.\
(\ref{bse1}). Finally, eq.\ (\ref{bse}) with the calculated kernel is used
to update the vertex function $\Phi$ (the $Y_i^m$). The new function $Y_1^0$
is used as described in sect.~3 to update $g_s$ and to test for convergence.

\subsection{Further Subroutines and Functions}

The subroutine {\bf int\_weights} calculates the integration points and
weights for  Gauss--Legendre
quadrature.

The RECURSIVE FUNCTION {\bf Chebyone} determines the values of the Chebychev
polynomials of the first kind.

The FUNCTION {\bf dq} implements the diquark function (\ref{V}).

\section{Testing the program}

Of course, trivial tests establishing the independence of the number of
integration points etc.\ have been performed. We also verified that the vertex
and wave functions are independent of the initializing functions.

In table \ref{conv} we demonstrate the convergence of the bound state 
mass in terms of $m_{\rm max}$ and $n_{\rm max}$. We have chosen parameter
set I from table \ref{pars} for that purpose. 
Since in the scalar diquark channel
the binding energy is just 46 MeV, one might not expect good convergence
due to the nearby poles in quark and diquark propagators. Nevertheless,
already for $m_{\rm max}=6$ and $n_{\rm max}=8$ the nucleon mass is exact
up to 0.1 MeV.

\begin{table}
 \begin{tabular}{llllllll} \hline \hline
   & $n_{\rm max}$ & 2 & 4 & 6 & 8 & 10 & 12 \\
 $m_{\rm max}$ \\ \hline
   0 &  & 0.891\hpt 0 & 0.915\hpt 6 & 0.921\hpt 2 & 0.922\hpt 5 & 0.922\hpt 8 & 0.922\hpt 9 \\
   2 &  & 0.919\hpt 3 & 0.932\hpt 1 & 0.935\hpt 2 & 0.936\hpt 0 & 0.936\hpt 2 & 0.936\hpt 3 \\
   4 &  &             & 0.936\hpt 5 & 0.938\hpt 1 & 0.938\hpt 6 & 0.938\hpt 7 & 0.938\hpt 8 \\
   6 &  &             &             & 0.938\hpt 7 & 0.938\hpt 9 & 0.939\hpt 0 & 0.939\hpt 0 \\
   8 &  &             &             &             & 0.939\hpt 0 & 0.939\hpt 0 & 0.939\hpt 0 \\ \hline \hline
 \end{tabular}
 \caption{Convergence of the bound state mass $M$ [GeV] in terms of 
  $m_{\rm max}$ and $n_{\rm max}$. Due to weaker convergence of
  the wave function, only $n_{\rm max}>m_{\rm max}$ is considered. 
  For momentum and angular integration $n_{|k|}=n_y=20$ grid points
  have been used.}
 \label{conv}
\end{table}

A very non-trivial test consists in changing the Mandelstam parameter $\eta$,
see the discussion the end of sect.~1. Please note that changing $\eta$ may
necessitate an increase in the order of the Chebychev expansion in order to
obtain  comparable accuracy. 

\section*{Acknowledgements}

We are grateful to Steven Ahlig, Christian Fischer, Gerhard Hellstern, 
Mike Pichowsky, Hugo Reinhardt, Craig Roberts, Sebastian Schmidt 
and Herbert Weigel for their collaboration on related issues in this model 
and helpful discussions.\\ 
We thank Sebastian Schmidt for a critical reading of this manuscript.\\
The work reported here has been supported by a Feodor--Lynen fellowship of
the Alexander-von-Humboldt foundation for M.O.\ and by COSY under
contract no.\ 4137660.\\
R.A.\ and L.v.S.\ want to express their gratitude to the members of the CSSM, 
University of Adelaide, for the hospitality experienced during their visits.

\newpage
\vskip 4cm

\section*{TEST RUN}

{\bf input file:}

\begin{scriptsize}
\begin{verbatim}
quark mass                            : 0.36
scalar diquark mass                   : 0.6255
axialvector diquark mass              : 0.684
diquark width                         : 0.95
type of diquark amplitude             : 1
exponent of diquark amplitude         : 2
momentum partitioning eta             : 0.36
npe  - momentum grid  |k|,|p|         : 20
ny   - angle grid \vec k \cdot \vec p : 20
mmax - Chebyshev accuracy VF          : 6
nmax - Chebyshev accuracy WF          : 8
accuracy for eigenvalue               : 1.D-9
accuracy for mass                     : 1.D-5
output file                           : sa.out

\end{verbatim}
\end{scriptsize}
\vskip 2cm

{\bf standard output:}
\begin{scriptsize}
\begin{verbatim}
 
 gs :     9.289520 ga :     6.965978
 
mass iteration #  1
 M_N =   0.684000
calculating H_kernel...
generated VF Chebyshev moments
 0  1  2  3  4  5  6 H_kernel done!
# it       1/sqrt(lambda)
   5 0.15781675145113D+02
  10 0.16128853436214D+02
  15 0.16125685785706D+02
  20 0.16125712415957D+02
  25 0.16125712185873D+02
  25 0.16125712185873D+02
|gs-1/sqrt(lambda)| / gs:   0.735904
 
mass iteration #  2
 M_N =   0.976984
calculating H_kernel...
generated VF Chebyshev moments
 0  1  2  3  4  5  6 H_kernel done!
# it       1/sqrt(lambda)
   5 0.66709436074472D+01
  10 0.71005125223069D+01
  15 0.70958216340534D+01
  20 0.70958710453183D+01
  25 0.70958705246610D+01
  27 0.70958705292115D+01
|gs-1/sqrt(lambda)| / gs:   0.236142
 
mass iteration #  3
 M_N =   0.905808
calculating H_kernel...
generated VF Chebyshev moments
 0  1  2  3  4  5  6 H_kernel done!
# it       1/sqrt(lambda)
   5 0.99728969409881D+01
  10 0.10508348360659D+02
  15 0.10500476420175D+02
  20 0.10500595152654D+02
  25 0.10500593361326D+02
  29 0.10500593387021D+02
|gs-1/sqrt(lambda)| / gs:   0.130370
 
mass iteration #  4
 M_N =   0.936576
calculating H_kernel...
generated VF Chebyshev moments
 0  1  2  3  4  5  6 H_kernel done!
# it       1/sqrt(lambda)
   5 0.88748406217450D+01
  10 0.93909382462111D+01
  15 0.93830314492941D+01
  20 0.93831520978449D+01
  25 0.93831502564257D+01
  29 0.93831502831329D+01
|gs-1/sqrt(lambda)| / gs:   0.010079
 
mass iteration #  5
 M_N =   0.938826
calculating H_kernel...
generated VF Chebyshev moments
 0  1  2  3  4  5  6 H_kernel done!
# it       1/sqrt(lambda)
   5 0.87863777879217D+01
  10 0.93005927655577D+01
  15 0.92927115567109D+01
  20 0.92928316413400D+01
  25 0.92928298111699D+01
  29 0.92928298376773D+01
|gs-1/sqrt(lambda)| / gs:   0.000356
 
mass iteration #  6
 M_N =   0.938908
calculating H_kernel...
generated VF Chebyshev moments
 0  1  2  3  4  5  6 H_kernel done!
# it       1/sqrt(lambda)
   5 0.87831354990374D+01
  10 0.92972802606202D+01
  15 0.92894000862408D+01
  20 0.92895201477319D+01
  25 0.92895183180267D+01
  29 0.92895183445258D+01
|gs-1/sqrt(lambda)| / gs:   0.000000
 
 done...
\end{verbatim}
\end{scriptsize}
\vskip 3cm

{\bf output file:}
\begin{scriptsize}
\begin{verbatim}
##################################################################
# mq       :  0.360000 ms        :  0.625500 ma        :  0.684000
# M        :  0.938908 ga/gs     :  0.749875 gs        :  9.289518
# eta      :  0.360000
# Dq.ampl. : RATIONAL  co        :  0.950000 nexp      :         2
# mmax     :         6 nmax      :         8
# npe      :        20 ny        :        20
##################################################################
#  VERTEX FUNCTION
#  S_1         |p|           0           1           2           3           4           5           6
        0.3448D-02  0.1000D+01  0.1038D-01  0.7696D-04  0.6986D-06  0.7131D-08  0.8341D-10  0.1139D-11
        0.1834D-01  0.9948D+00  0.5474D-01  0.2145D-02  0.1029D-03  0.5536D-05  0.3407D-06  0.2456D-07
        0.4590D-01  0.9678D+00  0.1305D+00  0.1237D-01  0.1432D-02  0.1842D-03  0.2677D-04  0.4692D-05
        0.8748D-01  0.8953D+00  0.2176D+00  0.3596D-01  0.7225D-02  0.1588D-02  0.3804D-03  0.1175D-03
        0.1453D+00  0.7662D+00  0.2796D+00  0.6502D-01  0.1845D-01  0.5651D-02  0.1777D-02  0.7879D-03
        0.2225D+00  0.5949D+00  0.2891D+00  0.8072D-01  0.2801D-01  0.1042D-01  0.3755D-02  0.2042D-02
        0.3237D+00  0.4110D+00  0.2470D+00  0.7298D-01  0.2822D-01  0.1145D-01  0.4387D-02  0.2579D-02
        0.4559D+00  0.2447D+00  0.1751D+00  0.4881D-01  0.2015D-01  0.8083D-02  0.3158D-02  0.1822D-02
        0.6289D+00  0.1197D+00  0.1011D+00  0.2350D-01  0.1047D-01  0.3758D-02  0.1465D-02  0.7767D-03
        0.8578D+00  0.4528D-01  0.4562D-01  0.7580D-02  0.4001D-02  0.1127D-02  0.4489D-03  0.2107D-03
        0.1166D+01  0.1244D-01  0.1527D-01  0.1415D-02  0.1123D-02  0.1986D-03  0.9469D-04  0.3774D-04
        0.1590D+01  0.2337D-02  0.3555D-02  0.8715D-04  0.2243D-03  0.1414D-04  0.1433D-04  0.4083D-05
        0.2193D+01  0.2812D-03  0.5378D-03 -0.1529D-04  0.2902D-04 -0.9603D-06  0.1489D-05  0.1644D-06
        0.3089D+01  0.1989D-04  0.4910D-04 -0.2865D-05  0.2078D-05 -0.1842D-06  0.8582D-07 -0.3406D-08
        0.4495D+01  0.7375D-06  0.2526D-05 -0.1597D-06  0.6870D-07 -0.6718D-08  0.1966D-08 -0.1851D-09
        0.6884D+01  0.1259D-07  0.6794D-07 -0.3476D-08  0.9234D-09 -0.7024D-10  0.1214D-10 -0.8321D-12
        0.1143D+02  0.8325D-10  0.7912D-09 -0.2673D-10  0.4159D-11 -0.2141D-12  0.1727D-13 -0.7142D-15
        0.2179D+02  0.1342D-12  0.2495D-11 -0.4574D-13  0.3731D-14 -0.9881D-16 -0.1453D-16 -0.1259D-18
        0.5451D+02  0.1406D-16  0.6590D-15 -0.4875D-17  0.1569D-18 -0.1834D-20  0.3594D-22 -0.4368D-24
        0.2901D+03  0.7739D-24  0.1932D-21 -0.2691D-24  0.1628D-26 -0.3582D-29  0.1522D-31 -0.4176D-34
#  S_2         |p|           0           1           2           3           4           5           6
        0.3448D-02 -0.2451D-02 -0.1331D-04 -0.1071D-06 -0.8120D-09 -0.5374D-11 -0.6284D-14  0.8165D-15
        0.1834D-01 -0.1299D-01 -0.3733D-03 -0.1594D-04 -0.6428D-06 -0.2298D-07 -0.2126D-09  0.8966D-10
        0.4590D-01 -0.3185D-01 -0.2217D-02 -0.2331D-03 -0.2338D-04 -0.2208D-05 -0.1132D-06  0.3099D-07
\end{verbatim}
\end{scriptsize}
\dots
\end{document}